\documentstyle{article}

\def\be{\begin{eqnarray}}
\def\ee{\end{eqnarray}}

\input{head.tex}

\def\theequation{\arabic{section}.\arabic{equation}}

\hoffset= - 1.0in         

\title{{\bf Resolving Puzzles of Massive Gravity
with and without violation
of Lorentz symmetry}
\vspace{.5cm}}
\author{{\bf Andrei Mironov}\footnote{ {\small {\it
Lebedev Physics Institute} and {\it ITEP, Moscow, Russia}};
mironov@itep.ru; mironov@lpi.ru}, {\bf Sergey Mironov}\footnote{
{\small {\it Moscow State University} and {\it ITEP, Moscow,
Russia}}; badzilla@rambler.ru}, {\bf Alexei
Morozov}\thanks{{\small {\it ITEP, Moscow, Russia}};
morozov@itep.ru}\ \ and\ {\bf Andrey Morozov}\thanks{{\small {\it
Moscow State University} and {\it ITEP, Moscow, Russia}};
Andrey.Morozov@itep.ru}\phantom{a}  }

\begin{document}

\maketitle

\vspace{-6.5cm}

\begin{center}
\hfill FIAN/TD-04/09\\
\hfill ITEP/TH-77/08
\end{center}

\vspace{4.5cm}

\begin{abstract}
\noindent We perform a systematic study
of various versions of massive gravity with and without
violation of Lorentz symmetry in arbitrary dimension.
These theories are well known to possess very unusual
properties, unfamiliar from studies of gauge and Lorentz
invariant models.  These peculiarities are caused by
mixing of familiar transverse fields with revived
longitudinal and pure gauge (Stueckelberg) fields
and are all seen already in quadratic approximation.
They are all associated with non-trivial dispersion laws,
which easily allow superluminal propagation, ghosts,
tachyons and essential irrationalities. Moreover, coefficients in
front of emerging modes are small, what makes the theories
essentially non-perturbative within a large Vainshtein radius.
Attempts to get rid of
unwanted degrees of freedom by giving them infinite masses
lead to DVZ discontinuities in parameter (moduli) space,
caused by un-permutability of different limits. Also, the condition
$m_{gh}=\infty$ can not be preserved already in non-trivial
gravitational backgrounds and is unstable under any other perturbations of
linearized gravity.
At the same time an {\it a priori} healthy model of massive
gravity in quadratic approximation definitely exists:
provided by any mass level of Kaluza-Klein tower.
It bypasses the problems because gravity field is mixed
with other fields, and this explains why such mixing helps
in other models. At the same time this can imply that the
really healthy massive gravity can still require infinite
number of extra fields beyond quadratic approximation.
\end{abstract}

\bigskip

\bigskip

\tableofcontents

\newpage

\section{Introduction}
\setcounter{equation}{0}

Renewed interest to massive gravity \cite{F,PF,magr,MGatt,othex,MT1}
and its further modifications,
involving mixing with extra light
fields and/or tiny violation of Lorentz symmetry
is dictated by problems of nowadays cosmology,
caused by spectacular advances of observation astronomy.
These days massive gravity is one of the so-far-desperate
attempts to cook up a theory which naturally explains
the phenomenology of hidden energy, which is currently
thought to account for over one half of the
energy density of the Universe.

Whatever is their relevance for phenomenological purposes,
the problems of massive gravity are
of deepest theoretical interest.
Abandoning Lorentz invariance one actually opens a Pandora
box of hidden structures, unobservable in the massless case.
They include a variety of dispersion (spectral) relations
for different components of the graviton field,
almost as rich as in solid state physics,
with non-quadratic dispersion laws, superluminal propagation,
emergency of non-trivial spatial structures (wave densities)
etc.

This paper (tightly connected with \cite{MT1}) is an attempt to understand in the most
primitive linear-algebra terms the puzzling properties
of massive gravity \cite{DVZ,Va,BD}
and of the whole new world arising after
the violation of Lorentz and gauge symmetries,
which was discovered in \cite{Ru,Du,TiT} and
nicely reviewed recently in \cite{RT}
(see also \cite{grapa}).
Since reasons for this strange behavior
are not the main concern for all these papers,
which are more interested in enumeration of different
models and their relevance cosmological applications,
our goal is to make a step in this direction.
Actually we are going to perform analysis in the
old-fashioned style of \cite{Ni} and of \cite{KM},
where the similar
puzzles of topological massive photodynamics \cite{3dph}
were addressed and resolved.
{\it A posteriori} it looks quite similar in spirit to the original
presentations in \cite{DVZ}, and also includes a direct generalization
of the original DVZ-approach to the case of Lorentz-non-invariant
theories and/or models with extra scalar fields.
It is of course very close to the original papers
\cite{Ru,Du,TiT,RT}, just our accents are different.
Our interest to the problem was initially motivated by studies of
massive graviton radiation \cite{New}-\cite{grarad}
in another class of speculative models,
related to {\it micro} rather than {\it macro} world:
in the TeV scale gravity \cite{TeVgr},
where masses are presumably of Kaluza-Klein origin and
various problems of the massive gravity are supposed to be absent
(other manifestations of the TeV scale gravity are discussed in \cite{mbh}-\cite{mtm}).

\bigskip

In the present paper
we analyze only quadratic approximation to
Einstein-Hilbert Lagrangian
(linearized gravity), thus all effects of field interactions,
including Vainshtein radius \cite{Va} or Boulware-Deser
modes \cite{BD} and superluminal effects \cite{OR}
in curved backgrounds are beyond the scope of this text.
As already explained in \cite{RT} they are in fact intimately related
to peculiar properties manifest in quadratic approximation,
in particular to the DVZ discontinuity \cite{DVZ} (see also \cite{Por}).

\bigskip

Given a quadratic action,
one immediately obtains the Born interaction
between currents:
\be
\phi_a K^{ab} \phi_b + {\cal J}^a\phi_a \ \ \
\longrightarrow  \ \ \
{\cal J}^a K^{-1}_{ab} {\cal J}^b =
\frac{{\cal J}^a\tilde K_{ab} {\cal J}^b}{\det K}
\ee
After Fourier transform
the entries of $K_{IJ}$ are quadratic polynomials in
the frequency $\omega$ and space momentum $\vec k$ (with some
mass terms added), and
\be
\det K = \prod_{a} \lambda_a(\omega,\vec k)
\ee
In the Lorentz-invariant case the $\vec k$-dependence is of course
related to $\omega$-dependence and
$\lambda_a(\omega ,\vec k) \rightarrow \lambda_a (k^2)$ with
$k^2 = -\omega^2+\vec k^2$ so that
\be\label{invdetK}
\frac{1}{\det K} = \sum_I \frac{A_a (\vec k)}{\omega^2-\Lambda_a(\vec k)}
\ee
however, Lorentz violation leads
to more sophisticated denominators in (\ref{invdetK}).
Coming back to the Born interaction, one can rewrite it as
\be\label{chan}
{\cal J}K^{-1}{\cal J} =
\sum_{a,b,c} \frac{\alpha^a_{bc}(\vec k){\cal J}^b{\cal J}^c}
{\lambda_a(\omega ,\vec k) + i\!\cdot\!0}
\ee
with some "structure constants" $\alpha(\vec k)$.

The problem of dispersion relations is basically that
of the eigenvalues of $K(k)$: roughly, $\omega = \varepsilon(|\vec
k|)$ is a condition that some eigenvalue $\lambda(k) = 0$. However,
this "obvious" statement requires a more accurate formulation. The
point is that $K$ is actually a quadratic form, not an operator,
what means that it can always be brought to the canonical form with
only $\pm 1$ and $0$ at diagonal, thus leaving no room to quantities
like $\lambda(k)$. Still, this "equally obvious" counter-statement
is also partly misleading, because we are interested not in an
isolated quadratic form, but in a family of those, defined over
${\cal M}$. This means that the sets of $\pm 1$ and $0$ can change
as we move along ${\cal M}$, and degeneracy degree of quadratic form
$K(k)$ can change. Of course, this degree (a number of $0$'s at
diagonal) is an integer and changes abruptly -- and thus is not a
very nice quantity. A desire to make it smooth brings us back a
concept of $\lambda(k)$. However, in order to introduce $\lambda(k)$
one needs an additional structure, for example, a metric in the
space of fields.

In application to our needs one can introduce "eigenvalues"
$\lambda(k)$ as follows: consider instead of
$\Pi = J\frac{1}{K}J$ a more general quantity
\be
\Pi(\lambda|k) = J\frac{1}{K-\lambda I}J
\ee
Then \textit{as a function of $\lambda$} it can be represented as a sum
of contributions of different poles:
\be
\Pi(\lambda|k) = \sum_{a,b,c}
\frac{\alpha_a^{bc} J_bJ_c}{\lambda_a-\lambda}
\ee
then $\lambda_a(k)$ are exactly the "eigenvalues"
that we are interested in, and our original
\be
\Pi(k) = \sum_{a,b,c}
\frac{\alpha_a^{bc}(k) J_b(-k)J_c(k)}{\lambda_a(k)}
\ee
The only thing that one should keep in mind is that this
decomposition depends on the choice of additional matrix
(metric) $I$, which can be chosen in different ways, in particular,
its normalization can in principle depend on the point of ${\cal M}$.
We shall actually assume that it does {\it not}, and clearly
the physical properties do not depend on this choice, however,
concrete expressions for $\lambda_a(k)$ do. It is important,
that the dispersion relations -- the zeroes of $\lambda_a(k)$ --
are independent of $I$.

Introduction of $I$ is also important from another point of view.
To be well-defined, the Lorentzian partition function
requires a distinction between the retarded and advanced correlators
(Green functions), which is usually introduced by adding
an infinitesimal imaginary term to the kinetic matrix $K$:
the celebrated $i\epsilon$ in the Feynman propagator.
However, in the case of kinetic \textit{matrix} this is not just $i\epsilon$,
it is rather $i\epsilon I_F$ with some particular matrix $I_F$. If
we identify our $I$ with $I_F$, then the dispersion relations are
actually
\be
\lambda_a(k) = i\epsilon
\label{iep}
\ee
what implies that
$\lambda_a(k)$ is, in fact, very different from $-\lambda_a(k)$, and
this is related to the important concept of {\it ghosts}.

When we have a family of theories, like massive gravities with
different masses, we actually have the set of quantities
$\lambda_a(\vec k)$ and $\alpha^a_{bc}(\vec k)$ "hanging"
over each point of parameter (moduli) space and one is
interested in the change of this structure when we move around
in the moduli space. What happens, different $\lambda_a$ can
cross or merge, they can also go away to infinity,
even more interesting are the properties of $\alpha$'s.
At some points of the moduli space the symmetry of the
underlying theory is enhanced, and we obtain a singularity,
where limits along different directions do not coincide
(this is exactly the reason for DVZ "discontinuity"),
so that such points should actually be
{\it blown up} to resolve the singularity.
All this is a typical string-theory subject \cite{UFN2},
it is amusing that this standard set of questions
unavoidably arises in the study of such seemingly innocent
subject as {\it linearized} massive gravity...

\bigskip

Schematically we consider different intermediate particles
(components of the gravity field), which contribute
in different channels in (\ref{chan}), or, putting this differently, just
diagonalize the coefficients $\alpha$ w.r.t. current indices
so that the sum (\ref{chan})
becomes a sum over channels and particles.
In general we are interested in a variety of channels
and in contribution of different species to each of this channel.
The whole pattern is characterized by the following data (see also \cite{MT1}):

{\bf Species:} dispersion laws. Sometimes the dispersion law is simple,
\be
\omega = \pm\sqrt{c^2\vec k^2+M^2}
\ee
however, the coefficients $c^2$ and $M^2$ are of importance.
The dispersion law with $c^2>1$ describe superluminals and with $M^2<0$ tachyons.
The superluminals always
travel faster than light
and can violate naive causality, \cite{Dsl},\cite{OR}. They are sometimes also
called tachyons in literature. But physically, they are very different from tachyons,
which are signals of instabilities and do not
violate causality (allow simultaneous but {\it uncorrelated and
causally independent} development of instabilities at space-like
intervals).
In fact we shall see that there are more sophisticated dispersion
laws
\be
\omega = \lambda(|\vec k|) \neq \pm\sqrt{c^2\vec k^2+M^2}
\ee
however their complexity does not increase with increase of the
dimension $d$: relations between $\omega$ and $|\vec k|$ are at most
quartic:
\be
\omega^2 = \sqrt{P_2(\vec k) \pm \sqrt{P_4(\vec k)}}
\ee
In the case of this more complicated dispersion law, we define $M^2$ as
position of the pole in $\omega^2$ at zero spatial momentum, and define
the tachyon mass square
as a real-valued solution to the equation $\Lambda(\vec k^2)=0$.

{\bf Residues:} coefficients $\alpha_a^{bc}$ controlling the contribution
of the specie $a$ to the channel $b$.
It is important to distinguish if ghost contributions
appear in physical (say, space-time transverse for conserved currents)
or unphysical channels (corresponding to sources of pure-gauge species).

{\bf Discontinuities} appear when some $M^2_a\rightarrow \infty$.
Accurate formulation of the problem
is that we look at the interaction
in a given channel at large distances, but not as large as $min(M_a^{-1})$,
so that all contributing species still look like massless.
However, if some $M_{a_0}=\infty$ there is simply no such region
and we observe a jump between "long-distance" interactions for
$M_{a_0} = \infty$ and $M_{a_0} = 0$.
In other words, switching on a {\it tiny} mass scale is not obligatory
a small effect if for some specie this {\it tiny} scale is multiplied
by infinity, constructed from other parameters (like $(A/B-1)^{-1}$ in
the Pauli-Fierz case below).

In the fully-comfortable theory there are no tachyons, ghosts and superluminals.
This, however, is rarely achievable in theories of massive spins,
greater than $1$, if one decides to abandon gauge invariance and
give masses by explicit rather than spontaneous breaking of gauge
invariance.

\bigskip

In the paper, we analyze the problem in a somewhat unusual way.
Instead of defining the normal modes by passing to Hamiltonian
formalism we directly diagonalize the kinetic matrices $K_{IJ}$
and $K_I^J$
in momentum representation. The
normal modes defined in the first way
are sometimes very convenient to deal with, especially in the case of massless
theories, since despite they are
introduced in explicitly Lorentz non-invariant way,
the longitudinal modes are very distinguished in this case,
 they carry a lot of physical information, in particular
easily distinguish between propagating and non-propagating
modes (without throwing the latter away from the spectrum as
one often does by imposing constraints in the Hamiltonian
formalism what in fact makes deformations to adjacent points
in the moduli space problematic, see \cite{KM} for initial
criticism of the standard approach).
On contrary, in the massive theories it is more convenient to use
the Lorentz invariant normal modes (the both types of
modes coincide in the rest frame).

In what follows we first start from a simplest example of electrodynamics
in order to illustrate the mode based approach and then continue in
s.\ref{LI} with massive and massless gravity.
One of our purposes in section \ref{LI} is
to justify the above-mentioned mechanism of DVZ discontinuity.
Then in s.\ref{LV}  we consider generalizations
to Lorentz-violated gravity, where the main novelty is
occurrence of quasiparticles with non-trivial dispersion laws,
not very familiar in elementary particle physics.
Finally in s.\ref{LS} we turn to massive gravity {\it mixed}
with some extra particles. Kaluza-Klein massive gravitons
belong to this class, what {\it a priori} explains why addition
of mixings {\it can} produce healthful theories of massive
gravity. For other fashionable models of this kind see
\cite{Du,MGatt}. The last section contains a discussion of various
physical consequences of the behaviour obtained in previous sections.

\newpage

\section{A warm-up example: massive photodynamics
\label{vec}}
\setcounter{equation}{0}

In this section we consider the case of photodynamics which is much simpler than the gravity and, hence,
we use it to illustrate the approach of the paper. We look at various patterns of adding masses,
including those breaking the Lorentz invariance.

\subsection{Generalities}

Photodynamics is the theory with quadratic action
\be
\int \Big(-\frac{1}{2}F_{\mu\nu}F^{\mu\nu} - M^2 A_\mu^2 + J_\mu A^\mu\Big) d^dx =
\int \Big(A^\mu(-k)K_{\mu\alpha}(k)A^\alpha(k) + A^{\mu}(-k)J_\mu(k)\Big)
d^dk
\label{MPAc}
\ee
Our immediate task is to:

-- enumerate different {\it modes}, contained in the field $A_\mu$,
which propagate through the space-time independently, without mixing,
and identify their properties,

-- find the sources of these modes,

-- decompose the Born interaction between the sources into contributions
of different modes.

In this way we can express various properties of interaction, mediated
by our theory, through the properties of individual modes and
in this way identify the origins of particular types of unusual behavior.

\bigskip

Of course, formal realization of this "program" is nothing but an
elementary linear algebra exercise with the kinetic matrix $K_{\mu\nu}(k)$
in momentum representation,
which in the case of photodynamics is simply an
ordinary symmetric $d\times d$ matrix:
\be\label{1}
K_{\mu\alpha} = k_\mu k_\alpha - (k^2+M^2)\eta_{\mu\alpha}
\ee
It can be easily diagonalized:
\be
K_{\mu\alpha} = \sum_{a=1}^d
\lambda_a v^{(a)}_\mu v^{(a)}_\alpha
\ee
where the $d$ eigenvectors are:
\be
\begin{array}{ccc}
{\rm gauging\ scalar} &
v_{g}^\mu 
= qk_\mu &
\\
{\rm longitudinal\ vector} & v_{||}^\mu &
\\
{\rm transverse\ vector}
& v_{i}^\mu, 
& i=1,\ldots,d-2
\end{array}
\ee
"Transverse" and "longitudinal" refer to {\it space}
rather than {\it space-time} vectors.
All these components are well defined at $M^2\neq 0$
and the splitting exhibits a smooth limit in the
massless limit $M^2\rightarrow 0$.

The gauge degree of freedom, $A_\mu = q k_\mu$
is scalar, it does not mix with the other
$d-1$ degrees of freedom,
\be
k^\mu K_{\mu\alpha} = -M^2k_\alpha
\ee
(it is a particular eigenvector of $K$),
but it has a non-trivial Lagrangian and even a kinetic
term for the {\it Stueckelberg field} $q(x)$, whenever
$M^2\neq 0$ and gauge invariance is broken.
The $d-1$ "physical" degrees of freedom in their turn
split into $1$+$(d-2)$  components -- longitudinal and
transverse photons with different eigenvalues
and different properties.

Note that the very definition of normal modes is not
Lorentz invariant (even if the theory is):
they solve an equation
$K_{\mu\alpha}v^\alpha = \lambda v^\mu$
and {\it not} $K_{\mu\alpha}v^\alpha = \tilde\lambda v_\mu$
This is important for making Lagrangian diagonal,
when expressed through the normal mode
-- this follows from orthogonality of matrix eigenvectors
(not that in the case of Euclidean signature
there would be no difference -- but the difference between
propagating and non-propagating modes is not seen there).

Born interaction,
\be
\int J^\mu(-k)P_{\mu\nu}(k)J^\nu(k) d^dk
\ee
is defined in terms of the propagator -- inverse of
kinetic matrix -- which is well defined for $M^2\neq 0$
\be
P_{\mu\alpha} = (K^{-1})_{\mu,\alpha} =
-\frac{\eta_{\mu\alpha} + \frac{k_\mu k_\alpha}{M^2}}{k^2+M^2}
\ee

\bigskip

Our {\bf notation} will be as follows:

Minkowski metric $\eta_{\mu\nu} = {\rm diag}(-1,+1,+1,\ldots,+1)$,

Space-time momentum $k_\mu = (\omega, k_{||}, 0,\ldots, 0)$,
$k^\mu = (-\omega, k_{||}, 0,\ldots,0)$,
spatial momentum will also be often denoted by $\vec k$ and
frequency $\omega$ -- by $k_0$.

The Lorentz square of space-time momentum
is $k^2 = -\omega^2 + \vec k^2 = -\omega^2 +  k_{||}^2$.

Space time indices are denoted by Greek characters:
$\mu,\nu =0,1,\ldots,d-1$, space indices -- by Latin characters
$i,j=1,\ldots,d-1$, transverse spatial indices
(in coordinate where $\vec k$ is directed along the first axis)
-- by $a,b=2,\ldots,d-1$.

\subsection{Massless photon
\label{mlph}}

This is the ordinary massless photodynamics.

{\bf Kinetic matrix} is:
\be
\left(\begin{array}{ccccc}
\omega^2 + k^2 & \omega k_{||} & 0 &  & 0\\
\omega k_{||} &  k_{||}^2 - k^2 & 0 &&0\\
0 & 0 & -k^2 & \ldots & 0 \\
&&\ldots &&\\
0&0&0&&-k^2
\end{array}\right) =
\left(\begin{array}{ccccc}
 k_{||}^2 & \omega k_{||} & 0&  & 0\\
\omega k_{||} & \omega^2 & 0 &  & 0\\
0 & 0 & \omega^2- k_{||}^2& \ldots & 0\\
&&\ldots &&\\
0&0&0&&\omega^2- k_{||}^2
\end{array}\right)
\ee

{\bf Eigenfunctions and eigenvalues:}
\be
\begin{array}{llll}
v_g^\mu  = &
\left(\begin{array}{c} -\omega \\  k_{||} \\ 0 \\ \ldots \\ 0
\end{array}\right)
= k^\mu & \lambda_g = 0 & {\rm gauging\ photon =
Stueckelberg\ scalar} \\
v_{||}^\mu = &
\left(\begin{array}{c}  k_{||} \\ \omega \\ 0 \\ \ldots \\ 0
\end{array}\right)
& \lambda_{||} = \omega^2+ k_{||}^2 \ \ \
&{\rm longitudinal\ photon} \\
v_{\bot}^\mu = &\
\underbrace{\left(\begin{array}{c} 0 \\ 0 \\ 1 \\ \ldots \\ 0
\end{array}\right),\ \ldots,\
\left(\begin{array}{c} 0 \\ 0 \\ 0 \\ \ldots \\ 1
\end{array}\right)}_{d-2 \ {\rm polarizations}} \ \ \
&\lambda_\bot = \omega^2- k_{||}^2 & {\rm transverse\ photon}
\end{array}
\ee
Thus $d-2$ transverse photons are just ordinary massless particles
with the normal kinetic term, described by the eigenvalue
$\omega^2- k_{||}^2$, which vanishes on-shell, when $\omega = k_{||}$.
This dispersion law allows non-vanishing values of $\omega$,
what implies that once emitted such particles can
{\it propagate} by "themselves".

Pure-gauge photon (gauging scalar or Stueckelberg scalar)
has vanishing eigenvalue, this means that it
completely drops out of the action: it "does not exist",
or is {\it ultralocal}: just equals to its source.

Longitudinal photon is a non-trivial field, but
the corresponding eigenvalue is positively defined
and vanishes (i.e. is {\it on shell}) at a single point
$\omega=k_{||}=0$: dispersion law is simply $\omega=0$.
This means that this field can not
exist "by itself", it is fully driven by the source.
Still, eigenvalue depends of $\omega$ and $\vec k$, what means
that even for a point-like source the field can be spread in
time and space.

{\bf Expanding} the gauge field $A_\mu$
in different sorts of photons,
\be
A^\mu(k) = A_g(k) v^\mu_g + A_{||}(k) v^\mu_{||}
+ A_\bot(k) v^\mu_{\bot}
\label{Avsv}
\ee
and substituting it into the action
$\int (F_{\mu\nu}F^{\mu\nu}+A^\mu J_\mu)$,
we obtain:
\be
A_{||}^2(\omega^2+k_{||}^2)^2 + A_\bot^2(\omega^2-k_{||}^2)
+ A_g(-\omega J_0+ k_{||}J_{||})
+ A_{||}(k_{||}J_0+\omega J_{||}) + A_\bot J_\bot
\ee
Note that we work with un-normalized eigenvectors: this
simplifies some formulas, though makes some other --
like this one for the Lagrangian -- look a little unusual.

The coupling $A_g J$ vanishes if $J_\mu$ is conserved,
in this case also the $A_{||} J$ coupling can be rewritten as
\be
A_{||}(k_{||}J_0+\omega J_{||})
= \frac{(\omega^2+k_{||}^2)}{\omega}A_{||} J_{||}
= \frac{(\omega^2+k_{||}^2)}{k_{||}}A_{||}J_{0}
\ee
so that the action, expressed in terms of the separated
variables, becomes
\be
A_{||}^2(\omega^2+k_{||}^2)^2 + A_\bot^2(\omega^2-k_{||}^2)
+ \frac{(\omega^2+k_{||}^2)}{\omega}A_{||} J_{||}
+ A_\bot J_\bot = \\ =
A_{||}^2(\omega^2+k_{||}^2)^2 + A_\bot^2(\omega^2-k_{||}^2)
+ \frac{(\omega^2+k_{||}^2)}{k_{||}}A_{||}J_{0}
+ A_\bot J_\bot
\ee

{\bf Born interaction} can be immediately read from this
formula:
\be
J_\alpha P^{\alpha\nu} J_\nu = 
{\omega^2- k_{||}^2} =
\frac{J_0^2}{k_{||}^2} + \frac{(J_\bot)^2}
{\omega^2- k_{||}^2} =
\frac{J_{||}^2}{\omega^2} + \frac{(J_\bot)^2}
{\omega^2- k_{||}^2}
\label{mlphJJ}
\ee
Of course it can be alternatively obtained by first writing
down the {\bf propagator} $P^{\alpha\nu}$ which satisfies
\be
K_{\mu\alpha}P^{\alpha\nu} = \delta^\nu_\mu - \frac{k_\mu k^\nu }{k^2}
\label{mlpheq}
\ee
and is equal to
\be
P^{\alpha\nu} = \frac{\eta^{\alpha\nu} + ck^\alpha k^\nu}
{\omega^2 - k_{||}^2}
\label{mlphpro}
\ee
so that the interaction of two conserved currents, such that
$k^\mu J_\mu = 0$, is
\be
J_\alpha P^{\alpha\nu} J_\nu = \frac{J_\alpha J^\alpha}
{\omega^2- k_{||}^2}
\ee
In order to obtain (\ref{mlphJJ}) from this simple expression
one should substitute an explicit resolution of conservation
constraint for the current:
\be
J_\alpha =
\left(\begin{array}{c}
J_0 \\ J_{||} \\ J_\bot
\end{array}\right) =
\left(\begin{array}{c}
ak_{||} \\ a\omega \\ J_\bot
\end{array}\right)
\ee
where $a=\frac{J_0}{k_{||}}=\frac{J_{||}}{\omega}$.

Note that because of the gauge invariance
(i.e. the vanishing of eigenvalue $\lambda_g$)
the propagator is not just inverse of kinetic matrix,
one should exclude the zero mode by putting transverse
matrix at the r.h.s. of (\ref{mlpheq}).
Instead propagator (\ref{mlphpro}) contains unspecified
coefficient $c$, which drops out from coupling to
conserved current.

\subsection{3d {\it topologically}
massive photodynamics \cite{3dph,KM}}

Consideration of the massive photodynamics we 
start with the celebrated "intermediate" example on the way from
massless to massive photodynamics, when photon gets a mass, but
gauge invariance is still unbroken and Stueckelberg fields do not
show up in the action. Despite propagating photon is massive there
is a pole at vanishing momentum in the interaction of currents.
Remarkably, this does not contradict unitarity because of existence
of a single propagating massless mode (not a particle) \cite{KM},
Since it is not a particle, the long-range interaction is {\it
topological}, namely in this case it is just an Aharonov-Bohm
interaction.

Lagrangian: \be -\frac{1}{2}F^{\mu\nu}F_{\mu\nu} +
M\epsilon^{\mu\nu\lambda}A_\mu F_{\nu\lambda} \ee

{\bf Kinetic $3\times 3$ matrix:} \be K_{\mu\nu} =
\left(\begin{array}{ccc}
\omega^2 + k^2 & \omega k_{||} & iM k_{||} \\
\omega k_{||} &  k_{||}^2 - k^2 & iM\omega \\
-iM k_{||} & -iM\omega & -k^2
\end{array}\right) =
\left(\begin{array}{ccc}
 k_{||}^2  & \omega k_{||} & iM k_{||} \\
\omega k_{||} & \omega^2 & iM\omega \\
-iM k_{||} & -iM\omega & \omega^2- k_{||}^2
\end{array}\right)
\ee The matrix is not fully symmetric, since the required symmetry
property is $K_{\mu\nu}(k)=K_{\nu\mu}(-k)$. Note also the appearance
of $i=\sqrt{-1}$ and that $\epsilon^{02}_{1} = \epsilon^{12}_0$
because of the Minkowski metric.

{\bf Eigenfunctions and eigenvalues:} \be
\begin{array}{llcl}
v_g^\mu  = & \left(\begin{array}{c} -\omega \\  k_{||} \\ 0
\end{array}\right)
= k^\mu & \lambda_g = 0 & {\rm 
Stueckelberg\ scalar} \\
v^\mu_+ = & \left(\begin{array}{c}  k_{||} \\ \omega \\ i\frac{
k_{||}^2-r}{M}
\end{array}\right)
& \lambda_{+} = \omega^2+ r &\ \ {\rm photon} \\
v^\mu_- = &\ \left(\begin{array}{c}  k_{||} \\ \omega \\ i\frac{
k_{||}^2+r}{M}
\end{array}\right)
& \lambda_- = \omega^2-r & {\rm \ photon}
\end{array} 
\ee where $r^2 \equiv  k_{||}^4 + M^2(\omega^2+ k_{||}^2)$.

Note that the {\bf eigenvalues $\omega^2 \pm r$ are quite
complicated}, however, the dispersion relation $\omega^2 = r$ is
equivalent to the standard one: \be \omega^2 =  k_{||}^2 + M^2 \ee
i.e. non-vanishing eigenvalues can be rewritten as \be \omega^2 \pm
r = \frac{(\omega^2- k_{||}^2 - M^2)(\omega^2+k_{||}^2)}{\omega^2\mp r}\ee so that no
irrationalities show up in denominators in the propagator \be
P_{\mu\alpha} = \frac{\eta_{\mu\nu} - \frac{k_\mu k_\nu}{k^2}}
{k^2-M^2} + \frac{iM\epsilon_{\mu\nu\lambda}k^\lambda}{k^2(k^2-M^2)}
\ee For conserved currents with $k_\mu J^\mu = 0$ \be J^\mu
P_{\mu\alpha} J^\alpha = \frac{1}{\omega^2-  k_{||}^2 - M^2}\left(
J_\bot^2 - \frac{2iMJ_\bot J_{||}}{\omega} +
\frac{k^2}{\omega^2}J_{||}^2\right) = \frac{1}{\omega^2-  k_{||}^2 -
M^2}\left( J_\bot^2 - \frac{2iMJ_\bot J_{0}}{ k_{||}} + \frac{k^2}{
k_{||}^2}J_{0}^2\right) \ee Note that {\bf inseparable products like
$k_{||}^2(k^2+M^2)$ appear in denominators}: this will be a typical
feature of all massive gauge theories, which does not depend -- as
we already see, on whether gauge invariance is preserved or not.

Not only the modes are inseparable, there is a pole at $p_{||}=0$,
despite particles are massive. This pole is indeed a {\bf long-range
interaction} and it is a question how such interaction can occur in
the theory of massive particles. The answer is \cite{KM} that the
spectrum of the theory is not exhausted by massive particles, there
is an additional single propagating mode without a gap. This is a
single mode, not a particle with the dispersion law $\omega = 0$
consistent with the Lorentz invariance (see more examples of such dispersion laws
in the gravity case with {\bf broken} Lorentz invariance, s.5, especially, s.5.6). 
Because of this the
long-interaction which it describes can not transfer space momentum
and is especially simple: it is a topological Aharonov-Bohm
interaction, but it exists, its effects are observable and it can
cause infrared divergencies in scattering cross-sections, just as
the ordinary long-range Coulomb interaction does in low
dimensions. 

As we saw, the irrationality, $r$ is gone from denominators in the
current-current interactions, moreover, it is not seen in the
numerators. This, however, is an illusion, irrationalities are
excluded at the price of considering non-diagonal interactions. {\bf If
the interaction is diagonalized}, i.e. written in terms of independent
modes (polarizations) $v^\mu$ and $\bar v^\mu$, 
{\bf the irrational $r$ is explicitly present} in
the formulas. This is again a standard feature of
massive gauge theories.

In fact, $v^\mu$ and $\bar v^\mu$ describe a left-polarized
massive photon with two massive degrees of freedom. There is an
extra degree of freedom (not a particle), responsible for the
long-range Coulomb interaction \cite{KM}.

\subsection{Massive photon}

This is the theory with the action (\ref{MPAc}).\footnote{
Since we wish to avoid higher derivatives for
Stueckelberg fields (not because they are bad,
higher derivatives theory suffers from ghosts at most,
and even this is not unavoidable \cite{AS}, see \cite{hder}
for a recent summary and a list of references,
but simply to somehow restrict our moduli space),
we do not consider another popular model of massive
photon, with $(\partial_\mu A^\mu)^2$ term
(ironically, this was exactly the model analyzed by
Stueckelberg in \cite{Stue}).
}
Here (and only here) we are going to consider for illustrative purposes the kinetic matrix
not as in (\ref{1}) with both lower indices, but with one lower and one upper. This leads to
Lorentzian eigenvalues, in contrast to the Euclidean ones obtained from (\ref{1}). This issue is
discussed in the next section, one can also find the comparison of the Euclidean and Lorentzian
eigenvalues in the massive photon case in \cite{MT1}.

Thus, now {\bf the Kinetic matrix} is
\be
\left(\begin{array}{ccc}
-\omega^2 - (k^2+M^2) & -\omega k_{||} & 0\\
\omega k_{||} &  k_{||}^2 - (k^2+M^2) & 0 \\
0 & 0 & -(k^2+M^2)
\end{array}\right) =
\left(\begin{array}{ccc}
 -k_{||}^2-M^2 & -\omega k_{||} & 0\\
\omega k_{||} & \omega^2 - M^2 & 0 \\
0 & 0 & \omega^2-( k_{||}^2+M^2)
\end{array}\right)
\label{MasPhotKM}
\ee
Its {\bf eigenvectors and eigenvalues} are
\be
\begin{array}{ccc}
v_g^\mu =
\left(
   \begin{array}{c}
     -\frac{\omega}{\sqrt{k_{||}^2-\omega^2}} \\
     \frac{k_{||}}{\sqrt{k_{||}^2-\omega^2}} \\
          0 \\
   \end{array}
 \right)\ \ \ \ \   &
 \lambda_g = - M^2
& {\rm the\ former\ Stueckelberg\ scalar}
\\ & \\
v_{||}^\mu =
\left(
   \begin{array}{c}
     -\frac{k_{||}}{\sqrt{k_{||}^2-\omega^2}} \\
     \frac{\omega}{\sqrt{k_{||}^2-\omega^2}} \\
     0 \\
   \end{array}
 \right)\ \ \ \ \  &
\lambda_{||}= \omega^{2}-k_{||}^{2}-M^2
&{\rm the\ former\ longitudinal\ photon}\atop{\rm (a\ scalar)}
\\ & \\
v_\bot^\mu =
\left(
   \begin{array}{c}
     0 \\
     0 \\
     1 \\
   \end{array}
 \right) &
\lambda_\bot = \omega^{2}-k_{||}^{2}-M^2
&{\rm transverse}\ (d-2)-{\rm vector}
\label{MasPhotEE}
\end{array}
\ee
Normalization of modes
is such that they polynomial in $\omega$ and $\vec k$.

\noindent
Gauge field $A_{\mu}$ is expanded in different sorts of photons
\be
A^{\mu}=A_{g}v^{\mu}_{g}+A_{||}v^{\mu}_{||}+A_{\bot}v^{\mu}_{\bot}
\label{MasPhotAE}
\ee
Lagrangian, when expressed through the normal modes,
becomes diagonal:
\be
\nonumber
-M^2A_{g}^2 + (M^2-\omega^2+k_{||}^2)A_{||}^2+(\omega^2-k_{||}^2-M^2)A_{\bot}^2
+J_\bot A_{\perp}+\frac{J_{||}k_{||}-J_0\omega}{\sqrt{k_{||}^2-\omega^2}}A_g+
\frac{-J_0k_{||}+J_{||}\omega}{\sqrt{k_{||}^2-\omega^2}}A_{||}
\label{MasPhotELE}
\ee

\bigskip

\noindent
Accordingly the {\bf Born interaction} of currents is
\be
-\frac{1}{4}\frac{M^2(J_0^2-J_{||}^2-J_{\bot}^2)-(J_0\omega+J_{||}k_{||})^2}{M^2(M^2+k_{||}^2-\omega^2)}
\label{MasPhotBI}
\ee
If, despite gauge invariance is broken, one keeps the currents
conserved, $k_\mu J^\mu = 0$, then the
Born interaction converts into
\be
-\frac{1}{4}\Big(\frac{J_0^2(k_{||}-\omega^2)}{k_{||}^2(M^2+k_{||}^2-\omega^2)}-
\frac{J_{\bot}^2}{(M^2+k_{||}^2-\omega^2)}\Big)=
-\frac{1}{4}\Big(\frac{J_{||}^2(k_{||}-\omega^2)}{\omega^2(M^2+k_{||}^2-\omega^2)}-
\frac{J_{\bot}^2}{(M^2+k_{||}^2-\omega^2)}\Big)
\label{MasPhotBItr}
\ee
{\bf Propagator}, the inverse of kinetic matrix, is
\be
\left(\begin{array}{ccc}
 \frac{M^2-\omega^2}{(\omega^2-k_{||}^2-M^2)M^2}
 & -\frac{\omega k_{||}}{(\omega^2-k_{||}^2-M^2)M^2} & 0\\
\frac{\omega k_{||}}{(\omega^2-k_{||}^2-M^2)M^2}
& -\frac{k_{||}^2+M^2}{(\omega^2-k_{||}^2-M^2)M^2}& 0 \\
0 & 0 & \frac{1}{\omega^2-(k_{||}^2+M^2)}
\end{array}\right)
\label{MasPhotpro}
\ee
and this provides for the Born interaction of currents:
\be
-\frac{1}{4}\Big(\frac{J_0^2(k_{||}-\omega^2)}{k_{||}^2(M^2+k_{||}^2-\omega^2)}-
\frac{J_{\bot}^2}{(M^2+k_{||}^2-\omega^2)}\Big)=
-\frac{1}{4}\Big(\frac{J_{||}^2(k_{||}-\omega^2)}{\omega^2(M^2+k_{||}^2-\omega^2)}-
\frac{J_{\bot}^2}{(M^2+k_{||}^2-\omega^2)}\Big)
\label{MasPhotproBI}
\ee
what coincides with (\ref{MasPhotBI}) and (\ref{MasPhotBItr}).
The poles at $\omega = 0$ and $\vec k = 0$ are spurious, as we already
discussed in s.\ref{mlph}. Moreover, in this case there is no
pole at $\omega=\vec k = 0$, and thus no long-range interactions 
as well, as can be seen from explicitly Lorentz-invariant formula at
the l.h.s. of (\ref{MasPhotproBI}).

Define the propagator as a solution to
$K_{\mu\alpha}P^{\alpha\nu} = \delta_\mu^\nu -
(1-\alpha)\frac{k_\mu k^\nu}{k^2}$,
$k^2 = -\omega^2+k_{||}^2$. Then, the smooth matching with
the massless case (\ref{mlpheq}) at $\alpha = 0$ is provided by the expression for the propagator
\be
P^{\mu\nu}=
\left(
  \begin{array}{ccc}
     \frac{k^{2}M^2-2{M^2}k_{0}^{2}+2k_{0}^{4}}
     {(-k_{0}^{2}+k^{2})M^2(-k_{0}^{2}+k^{2}+M^2)}
     & \frac{(M^2-2k_{0}^{2})k_{0}k}
     {(-k_{0}^{2}+k^{2})M^2(-k_{0}^{2}+k^{2}+M^2)} & 0 \\
  -\frac{k_{0}k(M^2+2k_{0}^{2})}{(-k_{0}^{2}+k^{2})M^2(-k_{0}^{2}+k^{2}+M^2)}
  &  \frac{k_{0}^{2}(2k^{2}+M^2)}{(-k_{0}^{2}+k^{2})M^2(-k_{0}^{2}+k^{2}+M^2)}
  & 0\\
  0 & 0 & -\frac{1}{-k_{0}^{2}+k^{2}+M^2} \\
  \end{array}
\right)
\ee

\subsection{Breaking Lorentz invariance}

We assume that the Lorentz invariance $SO(d-1,1)$
is broken down to the space-rotation invariance $SO(d-1)$
only by mass terms in the Lagrangian.
This means that instead of a single mass term
$M^2A_\mu^2 = M^2(-A_0^2+A_i^2)$
there can be two, $-m_0^2A_0^2 + m_1^2A_i^2$,
and $m_0$ does not necessarily coincide with $m_1$.
If gauge breaking terms were added to the kinetic terms
also, this would immediately provide $k^4$ terms
for the Stueckelberg fields in the Lagrangian.
Without gauge breaking one could only write
electric and magnetic terms $F_{0i}^2$ and $F_{ij}^2$
with different coefficients, but this is equivalent to
time rescaling and does not really break the Lorentz
invariance.

So, {\bf the kinetic matrix} is
\be
\left(\begin{array}{ccc}
\omega^2 + (k^2+M_0^2) & \omega k_{||} & 0\\
\omega k_{||} &  k_{||}^2 - (k^2+M_1^2) & 0 \\
0 & 0 & -(k^2+M_1^2)
\end{array}\right) =
\left(\begin{array}{ccc}
 k_{||}^2+M_0^2 & \omega k_{||} & 0\\
\omega k_{||} & \omega^2 - M_1^2 & 0 \\
0 & 0 & \omega^2-( k_{||}^2+M_1^2)
\end{array}\right)
\ee
Its {\bf eigenvectors and eigenvalues} are
\be
\begin{array}{cc}
\left(
   \begin{array}{c}
     1 \\
     -\frac{k_{||}^{2}-\omega^{2}+M_0^2+M_1^2 + r}{2\omega k_{||}} \\
     0 \\
   \end{array}
 \right)\ \ \ \ \  &
\lambda_-= \frac{1}{2}(\omega^{2} + k_{||}^{2}+M_0^2-M_1^2-r)
\\ & \\
\left(
   \begin{array}{c}
     1 \\
     -\frac{k_{||}^{2}-\omega^{2}+M_1^2+M_2^2 - r}{2\omega k_{||}} \\
          0 \\
   \end{array}
 \right)\ \ \ \ \   &
 \lambda_+ = \frac{1}{2}(\omega^{2}+ k_{||}^{2}+M_0^2-M_1^2 + r)
\\ & \\
\left(
   \begin{array}{c}
     0 \\
     0 \\
     1 \\
   \end{array}
 \right) &
\omega^{2}-k_{||}^{2}-M_1^2
\label{lvmvEE}
\end{array}
\ee
with
\be
r^2 = (\omega^2 + k_{||}^2 )^2
-2(M_0^2+M_1^2)(\omega^{2}-k_{||}^{2})+(M_0^2+M_1^2)^2
\ee
The gauge field $A_{\mu}$ is expanded in different sort of photons:

\be
A^{\mu}:=A_{-}v^{\mu}_{-}+A_{+}v^{\mu}_{+}+A_{\perp}v^{\mu}_{\perp}
\label{lvmvAE}
\ee
The Lagrangian, expressed through the normal modes
is diagonal:

\be
\nonumber
(P-Qr)A_{-}^2 +
(P+Qr)A_{+}^2
+(\omega^2-k_{||}^2-M_1^2)A_\bot^2 +J_\bot A_{\bot}+
\label{lvmvELE}
\ee \vspace{-0.2cm}
$$
+
\left(2J_0\omega k_{||}+J_{||}\omega^2
-J_{||}k_{||}^2-J_{||}(M_0^2+M_1^2)-J_{||}r\right)
A_{-}
+
\left(2J_0\omega k_{||}+J_{||}\omega^2
-J_{||}k_{||}^2-J_{||}(M_0^2+M_1^2)+J_{||}r\right)
A_{+}
$$
with
\be
P = (\omega^2-M_1^2)r^2, \ \ \ \
Q = M_1^2k_{||}^2-(2M_1^2+M_0^2)\omega^2+\omega^4+\omega^2k_{||}^2+M_0^2M_1^2+M_1^4
\ee

\smallskip

\noindent
and $\ r^2 = (\omega^2 + k_{||}^2 )^2-2(M_0^2+M_1^2)(\omega^{2}-k_{||}^{2})+(M_0^2+M_1^2)^2$,
so that $\ P^2-Q^2r^2 = 4\omega^2 k_{||}^2(\omega^2M_0^2-k_{||}^2M_1^2-M_0^2M_1^2)r^2$.

\bigskip

\noindent
{\bf The Born interaction} of currents is
\be
\frac{J_{\bot}^2}{\omega^2-k_{||}^2-M_1^2} +
\frac{\Big(2\omega k_{||} J_0 + (\omega^2-k_{||}^2-M_0^2-M_1^2-r)J_{||}\Big)^2}
{P-Qr} +
\frac{\Big(2\omega k_{||} J_0 + (\omega^2-k_{||}^2-M_0^2-M_1^2+r)J_{||}\Big)^2}
{P+Qr} 
\label{lvmvBI}
\ee
If the currents are conserved, this converts into
\be
\frac{J_{\bot}^2}{\omega^2-k_{||}^2-M_1^2}
- \frac{J_0^2(M_0^2\omega^2 - M_1^2k_{||}^2)}{(M_0^2\omega^2-M_1^2k_{||}^2-M_0^2M_1^2)k_{||}^2} =
\frac{J_{\bot}^2}{\omega^2-k_{||}^2-M_1^2}
- \frac{J_{||}^2(M_0^2\omega^2 - M_1^2k_{||}^2)}{(M_0^2\omega^2-M_1^2k_{||}^2-M_0^2M_1^2)\omega^2}
\label{lvmvBItr}
\ee
{\bf The propagator}, inverse of the kinetic matrix, is
\be
\left(\begin{array}{ccc}
 \frac{\omega^2-M_1^2}{(M_0^2\omega^2-M_1^2k_{||}^2-M^2)}
 & -\frac{\omega k_{||}}{(M_0^2\omega^2-M_1^2k_{||}^2-M^2)} & 0\\
-\frac{\omega k_{||}}{(M_0^2\omega^2-M_1^2k_{||}^2-M^2)}
& \frac{k_{||}^2+M_0^2}{(M_0^2\omega^2-M_1^2k_{||}^2-M^2)}& 0 \\
0 & 0 & \frac{1}{\omega^2-(k_{||}^2+M_1^2)}
\end{array}\right)
\label{lvmvpro}
\ee
and this provides for the Born interaction of currents:
\be
\frac{J_{\bot}^2}{\omega^2-k_{||}^2-M_1^2}
- \frac{J_0^2(M_0^2\omega^2 - M_1^2k_{||}^2)}{(M_0^2\omega^2-M_1^2k_{||}^2-M_0^2M_1^2)k_{||}^2} =
\frac{J_{\bot}^2}{\omega^2-k_{||}^2-M_1^2}
- \frac{J_{||}^2(M_0^2\omega^2 - M_1^2k_{||}^2)}{(M_0^2\omega^2-M_1^2k_{||}^2-M_0^2M_1^2)\omega^2}
\label{lvmvproBI}
\ee
what coincides with (\ref{lvmvBI}) and (\ref{lvmvBItr}).

\newpage

\section{Comment on the definition of normal modes}

It is a good moment now to illustrate our view  on
the normal modes \cite{MT1}.
We see that the physically relevant interaction (\ref{mlphJJ})
contains two very different kinds of structures.
In transverse channel interaction is clearly mediated by a
massless photon: there is a pole whenever this photon is
on-shell, $\omega^2=\vec k^2$.
At the same time in the temporal-longitudinal channel
the mediator is something very different:
the pole of the corresponding propagator is at $\omega=\vec k =0$
and nowhere else, i.e. has (real) codimension two.
The seeming codimension-one pole at $\omega = 0$ is spurious:
as $\vec k \rightarrow 0$ at $\omega \neq 0$ the $J_0$
in the numerator simultaneously tends to zero, so that
the would-be pole is fully eliminated.
Seemingly there is no pole at $\omega = 0$ but $\vec k\neq 0$.
These properties of the interaction are perfectly encoded
in the eigenvalues $\lambda$: $\lambda_\bot = 0$ exactly
on the mass shell of transverse photon, while $\lambda_{||}=0$
only at $\omega = \vec k = 0$.
This is why we consider {\it such} definition of $\lambda$'s
physically relevant.

However {\it such} definition may look somewhat non-conventional
and in fact inconvenient for other purposes.
In particular it is based on
an explicitly Lorentz-non-invariant definition of eigenvectors.
An alternative Lorentz-invariant definition,
however, provides another value of
$\tilde \lambda_{||} = \omega^2-\vec k^2$
and can be made consistent with the result for Born interaction
only within a sophisticated concept of a
"propagating, but decoupling" longitudinal photon.
Our approach rather treats longitudinal photon as totally
"non-propagating" -- in perfect accordance with (\ref{mlphJJ}).

Technically the difference is as follows.
We consider $K_{\mu\nu}$ as an ordinary symmetric matrix
and formally diagonalize it by orthogonal transformations,
i.e. by raising indices with the help of Euclidean
$\delta^{\mu\nu}$, instead of $\eta^{\mu\nu}$.
In other words we diagonalize the symmetric matrix
$$K_{\mu\nu}=\left(\begin{array}{cc} k_{||}^2 & \omega k_{||} \\
\omega k_{||} & \omega^2 \end{array}\right)$$ instead of
the asymmetric one $$K^\mu_\nu =
\left(\begin{array}{cc} -k_{||}^2 & -\omega k_{||} \\
\omega k_{||} & \omega^2 \end{array}\right)$$
The pure gauge (gauging) mode with the vanishing eigenvalue
is of course the same in both cases,
$ v^\mu_g = w^\mu_g =
\left(\begin{array}{c} -\omega \\ k_{||} \end{array}\right)$,
however for the longitudinal mode one gets
$v^\mu_{||} =
\left(\begin{array}{c} k_{||} \\ \omega \end{array}\right)$
with $\lambda_{||} = \omega^2+k_{||}^2$ instead of the usual
$w^\mu_{||} =
\left(\begin{array}{c} -k_{||} \\ \omega \end{array}\right)$
with $\lambda_{||} = \omega^2-k_{||}^2$.
Therefore, our non-covariant modes are orthogonal
in the ordinary linear-algebra sense,
i.e. w.r.t. Euclidean metric $\delta_{\mu\nu}$
instead of Minkowskian $\eta_{\mu\nu}$, while the Lorentzian
eigenmodes $w^\mu$ are orthogonal w.r.t. $\eta_{\mu\nu}$, i.e.
w.r.t. the group $SO(d-1,1)$ instead of $SO(d)$.
Accordingly,
the quadratic-form kinetic matrix is
$$K_{\mu\nu} = v_{||}^\mu v_{||}^\nu,\ \ \ \ \ 
\left(\begin{array}{cc} k_{||}^2 & \omega k_{||} \\
\omega k_{||} & \omega^2 \end{array}\right) =
\left(\begin{array}{c} k_{||} \\ \omega \end{array}\right)
\otimes \Big(k_{||}\ \ \omega\Big)$$ in terms of our
Euclidean normal mode, and our two eigenmodes,
used in (\ref{Avsv}), are just
$\delta^{\mu\nu}v^{||}_\nu$ and $\epsilon^{\mu\nu}v^{||}_\nu$.
Because of the different $\lambda_{||}$ our normal mode is
clearly non-propagating, while conventional longitudinal
photon does propagate, just "decouples".
As we shall see in the following
sections, when mass is introduced, gauge invariance broken
and the gauging (Stueckelberg) mode revived,
the difference gets even more pronounced:
our normal modes are the mixtures of Stueckelberg and
longitudinal modes, one propagating another non-propagating,
but both with non-trivial eigenvalues (\ref{MasPhotEE}),
while in the
Lorentz-covariant definitions Stueckelberg mode remains
non-propagating with the eigenvalue $M^2$, but longitudinal
photon remains propagating, acquires the standard dispersion
relation $\omega^2 = k_{||}^2+M^2$ -- the same as transverse
modes -- just it "no longer decouples". Thus the standard
view is not very helpful in visualizing the actual properties
of massive (gauge violating) theory with its sophisticated
propagators and Born interactions.

There is a clear physical reason to deal with the Euclidean eigenvalues
(see also \cite{MT1}).
Indeed, let us consider the massive 4-vector field Lagrangian
\be
-A_{\mu}(k^2+m^2)A^{\mu}=A^{\mu}K_{\mu\nu}A^{\nu}
\ee
It is ill-defined as contains ghosts, since the time component of the
field has the wrong sign of the time-derivative term. This is immediately reflected
in the corresponding negative derivative $\partial\lambda(\omega^2)/\partial\omega^2$
of one of the Euclidean eigenvalues of $K_{\mu\nu}$, while the
Lorentz eigenvalues are all positive in this case.

To put it differently, one takes care of the sign in the quadratic action when performing the 
Gaussian integration in the path integral. In particular, integrating over $A_{\mu}$, one's
concern is only on the coefficients in front of $A_\mu^2$, which are exactly 
the diagonal elements of the
Euclidean kinetic operator.

It deserves emphasizing that this subtle difference between Euclidean and Lorentz
eigenvalues is inessential when dealing with only the scalar sector
(which is of our main interest throughout the paper). Moreover, the majority of results below
are independent of this difference between Lorentz-invariant
and non-invariant definition of normal modes. In particular,
independent are $\det K$, characteristic equation and its
decompositions. Hence, dispersion laws, tachyons etc are independent of the choice
of modes, and only the ghost content could depend. More comments on this issue can be
found in \cite{MT1}, especially in Appendices I-II.

Note that
Lorentz covariant modes are easily restored from our formulas
below: in the rest frame (i.e. at $\vec k = 0$) the two
choices coincide and Lorentz transformation can be used
to obtain the Lorentz covariant modes in all other frames.
This is not so easy for non-Lorentz-covariant modes and we
have to write them down explicitly in all frames. However, even this advantage of the
Lorentz modes disappears when one works with non-Lorentz-invariant theories, as
we do in the second half of the paper.

In fact, in order to give the reader a flavour of difference between the Euclidean and
Lorentzian eigenmodes, for illustrative purposes, we present in a couple of places calculations for
the Lorentzian modes (as we did in s.2.4 above).

\newpage

\section{Lorentz invariant massive gravity
in quadratic approximation
\label{LI} }
\setcounter{equation}{0}

\subsection{Generalities}

Einstein-Hilbert action $\int R\sqrt{g}d^dx$
is highly non-linear in the metric field
and describes a pretty sophisticated interacting theory.
However, all the peculiarities of {\it massive} gravity
as compared to ordinary general relativity show up
already at the level of quadratic action for the
small deviation $h_{\mu\nu}$ of $g_{\mu\nu}$ from the
Minkowski background $\eta_{\mu\nu}$.
In this approximation it is also easy to introduce
a mass perturbation, which breaks gauge (general coordinate)
invariance and makes graviton field massive.
If Lorentz symmetry $SO(d-1,1)$ is preserved, the possible
deviations from General Relativity in quadratic approximation
are parameterized by two constants: $A$ and $B$.

Quadratic kinetic term \ \ \
$h^{\mu\nu}{\cal K}_{\mu\nu,\alpha\beta}h^{\alpha\beta}$\ \ is:
\be
{\cal K}_{\mu\nu,\alpha\beta} =
\Big( k_\mu k_\alpha\eta_{\beta\nu} +
k_\mu k_\beta\eta_{\alpha\nu} +
k_\nu k_\alpha\eta_{\beta\mu} +
k_\nu k_\beta\eta_{\alpha\mu}\Big) - \\
- 2\Big(k_\mu k_\nu\eta_{\alpha\beta} +
k_\alpha k_\beta \eta_{\mu\nu}\Big)
- (k^2 + A)\Big(\eta_{\mu\alpha}\eta_{\nu\beta} +
\eta_{\nu\alpha}\eta_{\mu\beta}\Big) +
2(k^2 + B)\eta_{\mu\nu}\eta_{\alpha\beta}
\label{LIgrK}
\ee

The propagator -- inverse of kinetic matrix -- can be
parameterized as following
\be\label{GF}
{\cal P}_{\mu\nu,\alpha\beta} =
a_1\Big( k_\mu k_\alpha\eta_{\beta\nu} +
k_\mu k_\beta\eta_{\alpha\nu} +
k_\nu k_\alpha\eta_{\beta\mu} +
k_\nu k_\beta\eta_{\alpha\mu}\Big)  +  \\  +
a_2 k_\mu k_\nu\eta_{\alpha\beta} +
a_3 k_\alpha k_\beta \eta_{\mu\nu}
+ a_4\Big(\eta_{\mu\alpha}\eta_{\nu\beta} +
\eta_{\nu\alpha}\eta_{\mu\beta}\Big) +
a_5\eta_{\mu\nu}\eta_{\alpha\beta} +
a_6k_\mu k_\nu k_\alpha k_\beta
\ee
and Born interaction of two stress-tensors is given by
\be\label{TT}
{\cal P}_{\mu\nu,\alpha\beta}T^{\mu\nu}T^{\alpha\beta} =
4a_1(kT)_\mu^2 + (a_2+a_3)(kTk)T + 2a_4T_{\mu\nu}^2 +
a_5T^2 + a_6(kTk)^2
\ \ \stackrel{k_\mu\!T^{\mu\nu}=0}{\longrightarrow}\ \
2a_4T_{\mu\nu}^2 + a_5T^2
\ee

Note that the Born
interaction of conserved energy-momentum tensors that involves only
$a_4$ and $a_5$ coefficients does not exhaust all
possible ways to observe gravity interactions -- one can just radiate
a graviton by the energy-momentum tensor, the radiated field being
\be\label{rad}
h_{\mu\nu}={\cal P}_{\mu\nu,\alpha\beta}T^{\alpha\beta} =
2a_1\left[k_\mu (kT)_\nu+k_\nu (kT)_\mu\right]+a_2k_\mu k_\nu T+
a_3 \eta_{\mu\nu}\left(kTk\right)+2a_4 T_{\mu\nu}+\\+a_5\eta_{\mu\nu} T
+a_6k_\mu k_\nu \left(kTk\right)
\ \ \stackrel{k_\mu\!T^{\mu\nu}=0}{\longrightarrow}\ \
a_2k_\mu k_\nu T+2a_4 T_{\mu\nu}+a_5\eta_{\mu\nu} T
\ee

Of course, violation of gauge invariance (general covariance)
in transition to massive gravity liberates stress-tensor from
obligation to be conserved.
However, this implies that violations are also present in
the matter sector.
If, as usual in the present-days discussions, we restrict all
violations of conventional physics to the pure gravity sector,
then the properties of matter are not changed and stress tensor
$T^{\mu\nu}$ remains conserved.
Technically, if one wants to break this property, a new
current, $J^\nu = k_\mu T^{\mu\nu}$, will appear in the
formulas, which should be further split into conserved
$\tilde J_\nu = J_\nu - k_\nu Q$ with $Q = \frac{1}{k^2}
k_\mu k_\nu T^{\mu\nu}$ and "improved", though non-local,
stress tensor
\be
\tilde T^{\mu\nu} = T^{\mu\nu} - \frac{1}{k^2}(k^\mu\tilde J^\nu
+ k^\nu\tilde J^\mu + k^\mu k^\nu Q)
\ee
will be conserved.

Therefore, (also in order to allow a smooth transition to the massless case) one
would better consider not an inverse of ${\cal K}$, but solution of
the equation ${\cal K}{\cal P} = E$ with the space-time transverse
r.h.s., explicitly accounting for the conservation of the
energy-momentum tensor, which is the source of the gravity field. To
make discussion complete, we introduce an additional parameter
$\alpha$ which interpolates between unity (at $\alpha =1$) and
space-time transverse (at $\alpha=0$) r.h.s. $E$: \be {\cal
P}_{\mu\nu,\rho\sigma}{\cal K}^{\rho\sigma}_{\alpha\beta} = \alpha
\left(\eta_{\mu\alpha}\eta_{\nu\beta}+
\eta_{\mu\beta}\eta_{\nu\alpha}\right)+(1-\alpha)
\left(\eta^t_{\mu\alpha}\eta^t_{\nu\beta}+
\eta^t_{\mu\beta}\eta^t_{\nu\alpha}\right)\ee where the transverse
Kronecker symbol is defined as \be \eta^t_{\mu\nu}\equiv
\eta_{\mu\nu}-{k_\mu k_\nu\over k^2} \ee

Then coefficients in the propagator are:

\bigskip

\be \hspace{-2.0cm}
\begin{array}{ccc}
a_1 =& -{1\over 2}{\alpha k^2 +\alpha A -A\over A k^2(k^2+A)}
\\
a_2=&{A-2B\over (k^2+A)\Big(A^2-dAB+(d-2)(B-A)k^2\Big)}
\\
a_3=&-{\left(B+\alpha B-A \right)k^2+AB(\alpha -1)\over
k^2(k^2+A)\Big(A^2-dAB+(d-2)(B-A)k^2\Big)} \\
a_4=&-{1\over 2}{1\over k^2+A}
\\
a_5=&-{(A-B)k^2+AB\over (k^2+A)\Big(A^2-dAB+(d-2)(B-A)k^2\Big)} \\
a_6=&{A^2(1-\alpha)(dB-A)+\alpha (d-2)(2B-A)k^4+(1-\alpha)
A\left[(d-3)A-(d-4)B \right]k^2\over Ak^4
(k^2+A)\Big(A^2-dAB+(d-2)(B-A)k^2\Big)}
\end{array}
\ee

\subsection{Normal modes of the gravity field}

After diagonalization of (\ref{LIgrK}) we get the following
decomposition of the gravity field $h_{\mu\nu}$:
\be
\frac{d(d+1)}{2}\ \ \  = \ \ \
\underbrace{\frac{(d-2)(d+1)}{2}}_{{\rm massive\
spin}\ 2}
 + \underbrace{\underbrace{(d-1)}_{{\rm space-time\ transverse}}
 +\underbrace{1}_{{\rm secondary}} }_{{\rm 
 Stueckelberg\ vector}} +
\underbrace{1}_{{\rm  space-time\ trace}} =
\label{degf0}
\ee
$$ =
\left\{
\underbrace{\frac{d(d-3)}{2}}_{{\rm spatial-transverse\ tensor}} +
\underbrace{(d-2)}_{{\rm longitudinal\ tensor}\atop{={\rm trasverse\ vector}}} +
\underbrace{1}_{{\rm spatial\ trace}}\right\} +
\left\{\underbrace{d-2}_{{{\rm spatial-transverse}}\atop
{{\rm Stueckelberg\ vector}}}
+ \underbrace{1}_{{\rm longitudinal}\atop{\rm Stueckelberg\ scalar}}
+ \underbrace{1}_{{\rm secondary}\atop{\rm Stueckelberg\ scalar}}\right\} + 1
$$
The first line here describes decomposition into
irreducible representations of $SO(d-1)$ in the rest frame,
while the second line -- that w.r.t. helicity group $SO(d-2)$
acting in the space orthogonal (space transverse) to the space
momentum $\vec k$.
In fact these decompositions remains relevant even if Lorentz
invariance is broken down to spatial-rotation symmetry $SO(d-1)$.
In gauge invariant theory Stueckelberg fields do not show up
in the Lagrangian (this also requires the sources to be transverse).
When gauge invariance is broken down by mass terms,
Stueckelberg fields acquire non-trivial kinetic terms and
they can mix with transverse degrees of freedom.

In arbitrary frame the two transverse vectors: one from
the massive graviton another Stueckelberg get mixed.
Also mixed are the four scalars.
This means that characteristic equation,
which defines eigenvalues of kinetic matrix, and is an
equation of degree $d(d+1)$ in $k_0=\omega$ and $\vec k$ is actually
factorized:
\be
{\rm Char}(\lambda) =
(\lambda-\lambda_{gr})^{\frac{d(d-3)}{2}}
P_2(\lambda)^{d-2}Q_4(\lambda) =
(\lambda-\lambda_{gr})^{\frac{d(d-3)}{2}}
(\lambda - \lambda_{vec}^+)^{d-2}(\lambda-\lambda_{vec}^-)^{d-2}
\prod_{a=1}^4(\lambda - \lambda_{sc}^a) = 0
\ee
where $P_2$ and $Q_4$ are polynomials of degree $2$ and $4$
respectively and all their coefficients as well as $\lambda_{gr}$
are quadratic functions of $\omega$ and $\vec k$.
In other words,

$\lambda_{gr}$ is some bilinear combination of $\omega$ and $\vec k$,

$\lambda_{vec}^{\pm} = p_2 \pm \sqrt{p_4}$, where $p_2$ and $p_4$
are respectively quadratic and quartic in $\omega$ and $\vec k$,

$\lambda_{sc}^{1,2,3,4}$ are the roots of degree-four polynomial.

\bigskip

In the rest frame the roots should be grouped in a different way:
\be
{\rm Char}(\lambda) = \left.(\lambda-\lambda_{gr})^{\frac{(d-2)(d+1)}{2}}
(\lambda - \lambda_{vec})^{d-1}(\lambda-\lambda_{sc}^+)
(\lambda-\lambda_{sc}^-)\right|_{\vec k = 0} = 0
\ee
i.e. at $\vec k = 0$
\be
\lambda_{vec}^+(\vec k=0) = \lambda_{gr}(\vec k=0),  \\
\lambda_{sc}^{spt}(\vec k=0) = \lambda_{gr}(\vec k=0),  \\
\lambda_{sc}^{spS}(\vec k = 0)=\lambda_{vec}^-(\vec k = 0)
\ee
where "spt" and "sSt" label spatial trace $h_{ii}$ and
spatial Stueckelberg scalar $h_{ij} = k_ik_js$ respectively.
The remaining two scalars, the space-time trace (stt) $h^\mu_\mu$
and secondary Stueckelberg scalar $h_{\mu\nu} = k_\mu k_\nu\sigma$
have eigenvalues, which are roots of quadratic equation:
\be
\left.\lambda^\pm_{sc} = q_2 \pm \sqrt{q_4}\right|_{\vec k = 0}
\label{rfqq}
\ee

\bigskip

\bigskip

In gauge invariant theory all the $d$ Stueckelberg fields have vanishing
eigenvalues and we get
\be
{\rm Char}(\lambda) = \left.\lambda^d
(\lambda-\lambda_{gr})^{\frac{d(d-3)}{2}}
(\lambda - \lambda_{vec}^+)^{d-2} (\lambda-\lambda_{sc}^{spt})
(\lambda-\lambda^{stt})\right|_{GI} = 0
\ee
i.e. in this case the two trace eigenvalues are roots
of quadratic equation,
\be
\lambda_{sc}^{\pm t} = \left.t_2 \pm \sqrt{t_4}\right|_{GI}
\label{GItt}
\ee
Actually gauge invariant will be only the massless
gravity (where, by the way, transition to the rest frame
is not a justified operation). The interrelation between modes in various cases
can be described by the following table:

\bigskip

\begin{tabular}{ccccc}
\underline{rest frame}&&\underline{normal modes}&&
\underline{gauge invariant (massless) case} \\
&&&&\\
&&graviton&$\rightarrow$& graviton\\
&$\swarrow$&&&\\
massive graviton &$\leftarrow\bigotimes$&&&\\
&$\nwarrow$&&&\\
&&vector&$\rightarrow$&vector\\
&&&&\\
&$\swarrow$&Stueckelberg vector&&\\
Stueckelberg $(d-1)$-vector&&&$\searrow$&\\
&$\nwarrow$&Stueckelberg scalar&$\rightarrow$&Stueckelberg $d$-vector\\
&&&$\nearrow$&\\
secondary Stueckelberg scalar&$\leftarrow$&secondary Stueckelberg scalar&&\\
&&&&\\
&$\bigotimes\leftarrow$&spatial trace&$\rightarrow$&spatial trace\\
&&&&\\
space-time trace&$\leftarrow$&space-time trace
&$\rightarrow$&space-time trace\\
\end{tabular}

\vspace{1cm}

It is also easy to describe the eigenvectors.
In coordinate system where $\vec k=k_1$ is directed along the first
axis the corresponding modes look as follows
(the matrix in the upper left corner is formed by  directions $0$ and $1$):
$$
\begin{array}{cc}
\left(\begin{array}{cc|ccc}
0 & 0 & 0 & & 0 \\
0 & 0 & 0 &\ldots & 0 \\ \hline
0 &  0 & && \\
& \ldots &&h_{ab}& \\
0 & 0 &&&
\end{array}\right) & {\rm transverse\ traceless\ graviton}
\\
\\
\left(\begin{array}{cc|ccc}
0 & 0 & \ldots & v & \ldots  \\
0 & 0 & \ldots & w & \ldots  \\
\hline
 & \ldots & && \\
 v & w & && \\
 & \ldots &&&
\end{array}\right) & {\rm two\  transverse}\ (d-2)-{\rm vectors}\ v\ \hbox{and}
\ w
\\
\\
\left(\begin{array}{cc|ccc}
\alpha & \beta & &\ldots & \\
\beta & \gamma & && \\ \hline
& & \delta && \\
\ldots & &\ldots & \\
&&&& \delta
\end{array}\right) & {\rm four\ scalars}\ \alpha, \beta, \gamma\ \hbox{and}\ \delta
\end{array}
$$
Again these shapes remain the same even after Lorentz symmetry
violation  $SO(d-1,1)\rightarrow SO(d-1)$.

\subsection{Massless gravity}

This is conventional General Relativity with $A=B=0$.

Kinetic matrix (\ref{LIgrK}) is actually a symmetric
matrix of size $\frac{d(d+1)}{2}$ when acting in the space of normal modes:

\bigskip

\begin{equation}
\left[\begin{array}{l|ccc|cc|ccc}
&00 & 01 & 11 & 0a & 1a & aa & ab & bb \\
\hline
00&&&&&&-k_{||}^2&&-k_{||}^2\\
01&&&&&&-\sqrt{2}\omega k_{||}&&-\sqrt{2}\omega k_{||}\\
11&&&&&&-\omega^2&&-\omega^2\\
\hline
0a&&&&k_{||}^2&\omega k_{||}&&&\\
1a&&&&\omega k_{||}&\omega^2&&&\\
\hline
aa&-k_{||}^2&-\sqrt{2}\omega k_{||}&-\omega^2&&&&&-\omega^2+k_{||}^2\\
ab&&&&&&&\omega^2-k_{||}^2&\\
bb&-k_{||}^2&-\sqrt{2}\omega k_{||}&-\omega^2&&&-\omega^2+k_{||}^2&&\\
\hline
\end{array}\right]
\label{mlgKM}
\end{equation}

\bigskip

where we denote it by square brackets, keeping ordinary
brackets for $d\times d$ matrices, like gravity field $h_{\mu\nu}$.

Its {\bf eigenvectors} ($d\times d$ matrices) {\bf and eigenvalues} are:
$$
\begin{array}{ccc}
\omega^2-k_{||}^2 &
\left(\begin{array}{cc|ccc}
0 & 0 & \ldots & 0 & \\
0 & 0 & & 0 & \\ \hline
&\ldots &&&\\
0 &  0 & & h_{ab} & \\
&\ldots &&&
\end{array}\right) & {\rm transverse\ traceless\ graviton} \\
\\
\omega^2 + k_{||}^2 &
\left(\begin{array}{cc|ccc}
0 & 0 & \ldots & k_{||} & \ldots \\
0 & 0 & &\omega &\ldots\\ \hline
& \ldots &&&\\
k_{||} &  \omega & & 0 & \\
& \ldots &&&
\end{array}\right) & {\rm transverse}\ (d-2)-{\rm vector} \\
\\
0 &
\left(\begin{array}{cc|ccc}
0 & 0 & \ldots & -\omega & \\
0 & 0 & & k_{||} & \\ \hline
& \ldots &&&\\
-\omega &  k_{||} & & 0 & \\
& \ldots &&&
\end{array}\right) & {\rm Stueckelberg}\ (d-2)-{\rm vector} \\
\\
0 &
\left(\begin{array}{cc|ccc}
 -\omega & k & \ldots & 0& \\
k & 0 & & 0 &\\ \hline
& \ldots &&& \\
0 &  0 & & 0& \\
& \ldots &&&
\end{array}\right)
\ \ {\rm or} \ \
\left(\begin{array}{cc|ccc}
0& -\omega  & \ldots & 0 & \\
-\omega & k  & & 0 & \\ \hline
& \ldots &&&\\
0 &  0 & &0 & \\
& \ldots &&&
\end{array}\right)
 & \displaystyle{{\rm temporal\ or\ spatial}\atop{\rm Stueckelberg\ scalar}}\\
\\
0 &
\left(\begin{array}{cc|ccc}
\omega^2 & 0 & \ldots & 0 & \\
0 & -k^2 & & 0 & \\ \hline
& \ldots &&&\\
0 &  0 & & 0 & \\
& \ldots &&&
\end{array}\right) & {\rm secondary\ Stueckelberg\ scalar}\\
\\
\lambda_{\pm} &
\left(\begin{array}{cc|ccc}
k^2 & \omega k & \ldots & 0 & \\
\omega k & \omega^2  & & 0 & \\ \hline
& \ldots & \lambda_{\pm} && \\
0 &  0 & & \lambda_\pm & \\
&\ldots &&& \lambda_{\pm}
\end{array}\right) & \displaystyle{{\rm two\ mixed\ traces,}\atop
{\rm spatial\ and\ space-time}} \\
\end{array}
\label{mlgEE}
$$
with
\be
\lambda_{\pm} = (d-3)(-\omega^2+\vec k^2) \pm r = t_2 \pm \sqrt{t_4},\\
r^2 = (d-1)^2\omega^4 -2(d^2-10d+17)\omega^2{\vec k}^2 + (d-1)^2\vec k^4
= (d-1)^2(\omega^2-\vec k^2)^2 + 16(d-2)\omega^2{\vec k}^2
\ee
In other words, in massless (gauge invariant) case
\be
\lambda_{gr} = \omega^2-\vec k^2,  \\
\lambda_{vec}^+ = \omega^2+\vec k^2,  \\
\lambda_{vec}^- = 0
\ee
and $Q_4(\lambda)$ factorizes, in accordance with (\ref{GItt}):
\be
Q_4(\lambda) = \lambda^2\Big(\lambda^2+(d-3)(\omega^2-\vec k^2)\lambda -
(d-2)(\omega^2+\vec k^2)^2\Big),
\ee
so that the two Stueckelberg scalars have $\lambda_{Sts}^\pm = 0$
and the two trace eigenvalues are
\be
\lambda^\pm_{tr} = \frac{1}{2}\left((d-3)(-\omega^2+\vec k^2)
\pm\sqrt{(d-1)^2(\omega^4+\vec k^4) - 2(d^2-10d+17)\omega^2\vec k^2}\right)
\ee

\subsection{Massive gravity in the rest frame}

Rest frame, where $k_{||}=0$ is distinguished because $SO(d-1)$ symmetry
is fully restored in it and there is no reason to distinguish between
$1$ and other spatial directions. Accordingly the kinetic matrix is:

\bigskip

\begin{equation}
\left[\begin{array}{l|cc|ccc}
&00 & 0i & ii & ij & jj \\
\hline
00&B-A&0&-B&0&-B\\
0i&0&A&0&0&0\\
\hline
ii&-B&0&B-A&0&-\omega^2+B\\
ij&0&0&0&\omega^2-A&0\\
jj&-B&0&-\omega^2+B&0&B-A\\
\hline
\end{array}\right]
\label{rfgKM}
\end{equation}

\bigskip

Its {\bf eigenvectors} ($d\times d$ matrices) {\bf and eigenvalues} are:
\begin{equation}
\begin{array}{ccc}
\omega^2 -A &\left(\begin{array}{cc|c}
0 & 0 & 0 \\
0 & 0 & 0 \\ \hline
0 &  0 & h_{ab}
\end{array}\right) & {\rm transverse\ traceless\ graviton} \\
\\
\omega^2 -A&\left(\begin{array}{cc|ccc}
0 & 0 & \dots&0&\ldots \\
0 & 0 & \ldots&1&\ldots \\ \hline
&\ldots\\
0 &  1 & &0\\
&\ldots
\end{array}\right) & {\rm transverse}\ (d-2)-{\rm vector} \\
\\
A&\left(\begin{array}{cc|ccc}
0 & 0 & \ldots&1&\ldots \\
0 & 0 & \ldots&0&\ldots\\ \hline
&\ldots\\
1 &  0 && 0\\
&\ldots
\end{array}\right) & {\rm Stueckelberg}\ (d-2)-{\rm vector} \\
\\
A&
\left(\begin{array}{cc|c}
0& 1  & 0 \\
1 & 0  & 0 \\ \hline
0 &  0 & 0
\end{array}\right)&
 {\rm Stueckelberg\ scalar}\\
\\
\omega^2-A&\left(\begin{array}{cc|c}
0 & 0 & 0 \\
0 & d-2 & 0 \\ \hline
0 &  0 & -1
\end{array}\right) & {\rm secondary\ Stueckelberg\ scalar}\\
\\
\lambda_{sc}^{\pm}&\left(\begin{array}{cc|c}
-(d-1)B & 0 & 0 \\
0 & \lambda_{sc}^{\pm}+A-B  & 0 \\ \hline
0 &  0 & \lambda_{sc}^\pm+A-B
\end{array}\right) & {\rm two\ mixed\ traces,\ spatial\ and\ space-time} \\
\end{array}
\end{equation}
or, if we do not distinguish between the $1$ and other directions,
\be
\begin{array}{ccc}
\omega^2-A&\left(\begin{array}{c|c}
0 & 0 \\ \hline
0 & h_{ij}
\end{array}\right) & {\rm  traceless\ graviton} \\
A&\left(\begin{array}{c|c}
0 & 1 \\ \hline
1 &  0
\end{array}\right) & {\rm Stueckelberg}\ (d-1)-{\rm vector} \\
\lambda_{sc}^\pm&\left(\begin{array}{c|c}
-(d-1)B & 0 \\ \hline
0& \lambda_{sc}^\pm+A-B
\end{array}\right) & {\rm
a\ mixture\ of\ secondary\ Stueckelberg\ scalar\
and\ space-time\ trace} \\
\end{array}
\ee
In other words, one has in the rest frame in the massive gravity case the
three possible values of the eigenvalues:
\be
\begin{array}{ccc}
& \lambda_{vec} = A & d-1\ {\rm times} \\
&\lambda_{gr} = \omega^2- A & \frac{(d+1)(d-2)}{2}\ {\rm times} \\
& \lambda_{sc}^\pm = {-(d-2)\omega^2 + dB-2A\
\pm\sqrt{(d-2)^2(B-\omega^2)^2 + 4(d-1)B^2}\over 2}
& {\rm two\ traces}
\end{array}
\ee
The masses present in the spectrum can be obtained by solving the equations
$\lambda_i (\omega^2=m^2)=0$. This gives the mass of the graviton multiplet $m^2=A$
and the mass of the two mixing scalars $M^2=\frac{A(dB-A)}{(d-2)(A-B)}$
(the equation $\lambda_{sc}^\pm=0$ has only one solution).
Note that at any values of $A$ and $B$, one of the
two eigenvalues that crosses the abscissa axis has the negative slope at the crossing point.
This means it is a ghost. One may say that, depending on relations between $A$ and $B$,
$M^2$ can be positive or negative, in the latter case the ghost do not propagate (and is
the tachyon, in fact). However, taking non-zero and large enough $\vec k^2$, one can get
a positive solution to the equation $\lambda_{sc}^\pm=0$ (due to the Lorentz invariance
only the invariant combination $\omega^2-\vec k^2$ enters the equation).

The only possibility to avoid the ghost is to put
$A=B$ (the celebrated Pauli-Fierz case), when the mass
of the ghost becomes infinite (the equation $\lambda_{sc}^\pm=0$ has no solutions at
all) and goes away from the spectrum.

Let us return now to the propagator. The coefficients in the propagator can be
rewritten as

\bigskip

\be
\begin{array}{ccc}
a_1 =&  \frac{1}{2m^2}\left(\frac{1-\alpha}{k^2} - \frac{1}{k^2+m^2}\right)\\
a_2=& \frac{1}{(d-1)m^2} \left(\frac{1}{k^2+m^2} - \frac{1}{k^2+M^2}\right)\\
a_3=& a_2 +
\frac{(1-\alpha)B}{A(dB-A)}\left(\frac{1}{k^2} - \frac{1}{k^2+M^2}\right)\\
a_4=& -\frac{1}{2}\cdot\frac{1}{k^2+m^2}\\
a_5=& \frac{1}{(d-1)(d-2)}
\left(\frac{d-2}{k^2+m^2} + \frac{1}{k^2+M^2}\right)\\
a_6=&{1\over m^4}{d-2\over d-1}\left({1\over k^2+M^2}-{1\over k^2+m^2}\right)
+{1-\alpha\over m^4}\left\{-{m^2\over k^4}+{(d-2)M^2+m^2\over M^2(d-1)}
\left({1\over k^2}-{1\over k^2+M^2}\right)
\right\}
\end{array}
\label{a16}
\ee

\bigskip

\bigskip

Here we again encounter two different dispersion laws: $k^2+m^2=0$ and
$k^2+M^2=0$ with \be m^2=A,\ \ \ \ \ M^2 =
\frac{A(dB-A)}{(d-2)(A-B)} \ee There is also a fictitious pole at
$k^2=0$ that contributes only at $\alpha\neq 1$ and makes the
transverse part of the propagator. Therefore, it cancels with the
conserved energy-momentum tensor.

As we already discussed the $M^2$ mode in some channels (coefficients $a_i$)
behaves as a ghost, i.e. enters with minus sign
as compared to the $m^2$ mode, but this does {\it not} happen in the
channels attached to conserved currents. It, however, enters the radiation,
(\ref{rad}) in the coefficient $a_2$. Note that this coefficient cancels
in the other distinguished case when the both masses are equal,
\be
m^2=M^2  \ \ \ \ {\rm when} \ \ \ \ A=2B
\ee

The receipt to remove the ghost we mentioned is to bring its mass to infinity.
In this Pauli-Fierz limit, the coefficients (\ref{a16}) becomes
\be
\hspace{-2.0cm}
\begin{array}{ccc}
a_1 =&  \frac{1}{2m^2}\left(\frac{1-\alpha}{k^2} - \frac{1}{k^2+m^2}\right)\\
a_2=& \frac{1}{(d-1)m^2} \frac{1}{k^2+m^2}\\
a_3=& a_2 +
\frac{(1-\alpha)}{(d-1)m^2}\frac{1}{k^2}\\
a_4=& -\frac{1}{2}\cdot\frac{1}{k^2+m^2}\\
a_5=& \frac{1}{(d-1)(k^2+m^2)} \\
a_6=&-{1\over m^4}{d-2\over d-1}{1\over k^2+m^2}
+{1-\alpha\over m^4}\left(-{m^2\over k^4}+{d-2\over d-1}
{1\over k^2}
\right)
\end{array}
\label{PF}
\ee
Interaction of two stress tensors looks in this case especially simple if
$\alpha=1$:
\be
-{1\over k^2+m^2}\left[(T_{\mu\nu})^2 - \frac{1}{d-1}(T^\lambda_\lambda)^2
+\frac{2}{m^2}t_\mu^2 -
\frac{2}{m^2(d-1)}(kt)T^\lambda_\lambda
+ \frac{d-2}{m^4(d-1)}(kt)^2\right]
\label{PFint}
\ee
where $k^\mu T_{\mu\nu} = t_\nu$.
For a conserved stress tensor $t=0$.

However, beyond the quadratic approximation ghosts show up through the
Boulware-Deser instability in curved backgrounds even in the Pauli-Fierz case
(it is enough to
consider quadratic perturbations but near the non-trivial metric
$g_{\mu\nu}=\rho \eta_{\mu\nu})$.

\subsection{Generic frame}

The kinetic matrix is now

\bigskip

\begin{equation}
\left[\begin{array}{l|ccc|cc|ccc}
&00 & 01 & 11 & 0a & 1a & aa & ab & bb \\
\hline
00&B-A&&-B&&&-k_{||}^2-B&&-k_{||}^2-B\\
01&&A&&&&-\sqrt{2}\omega k_{||}&&-\sqrt{2}\omega k_{||}\\
11&-B&&B-A&&&-\omega^2+B&&-\omega^2+B\\
\hline
0a&&&&k_{||}^2+A&\omega k_{||}&&&\\
1a&&&&\omega k_{||}&\omega^2-A&&&\\
\hline
aa&-k_{||}^2-B&-\sqrt{2}\omega k_{||}&-\omega^2+B&&
&B-A&&-\omega^2+k_{||}^2+B\\
ab&&&&&&&\omega^2-k_{||}^2-A&\\
bb&-k_{||}^2-B&-\sqrt{2}\omega k_{||}&-\omega^2+B&&
&-\omega^2+k_{||}^2+B&&B-A\\
\hline
\end{array}\right]
\label{LImgKM}
\end{equation}

\bigskip

Lorentz-invariant eigenmodes are much simpler: they are obtained by
Lorentz transformations from those in the rest frame.
Accordingly the eigenvalues are
\be
\begin{array}{ccc}
& \lambda_g = A & d-1\ {\rm times} \\
&\lambda_{gr} = \omega^2-\vec k^2 - A & \frac{(d+1)(d-2)}{2}\ {\rm times} \\
& \lambda_{sc}^\pm = {(d-2)k^2 + dB-2A\
\pm\sqrt{(d-2)^2(k^2+B)^2 + 4(d-1)B^2}\over 2}
& {\rm two\ traces}
\end{array}
\ee
and the corresponding eigenmodes $w^{\mu\nu}$ are
$$
\underbrace{\left(\begin{array}{cc|ccc}
0 & 0 &\ldots &-\omega &\ldots \\
0 & 0 & &k_{||} &\\ \hline
 & \ldots &&&\\
-\omega &k_{||} &&0 &\\
 & \ldots &&&
\end{array}\right)}_{d-2}, \ \ \ \ \
\left(\begin{array}{cc|ccc}
\omega k_{||} & -\frac{\omega^2+k_{||}^2}{2} &&\ldots & \\
-\frac{\omega^2+k_{||}^2}{2} & \omega k_{||} & & &\\ \hline
 &&&&\\
 &\ldots &&0 &\\
 & &&&
\end{array}\right)
$$
$$
\underbrace{
\left(\begin{array}{cc|ccc}
0 & 0 & &\ldots & \\
0 & 0 & &&\\ \hline
 & \ldots &&&\\
 &&& h_{ab} &\\
 & \ldots &&&
\end{array}\right)}_{\frac{d(d-3)}{2}}, \ \ \ \ \
\underbrace{\left(\begin{array}{cc|ccc}
0 & 0 &\ldots &-k_{||} &\ldots \\
0 & 0 & &\omega &\\ \hline
 & \ldots &&&\\
-k_{||} &\omega &&0 &\\
 & \ldots &&&
\end{array}\right)}_{d-2}, \ \ \ \ \
\left(\begin{array}{cc|ccc}
k_{||}^2 & -\omega k_{||} &&\ldots & \\
-\omega k_{||} & \omega^2 & &\omega &\\ \hline
 &  &\frac{k_{||}^2-\omega^2}{d-2}&&\\
&\ldots &&\frac{k_{||}^2-\omega^2}{d-2} &\\
 & &&&\frac{k_{||}^2-\omega^2}{d-2}
\end{array}\right)
 $$
 $$
\left(\begin{array}{cc|ccc}
\lambda^\pm_{sc}+A+(d-2)\omega^2 & \sqrt{2}(d-2)\omega k_{||} &&\ldots & \\
\sqrt{2}(d-2)\omega k_{||} & -\lambda^\pm_{sc}-A+(d-2)k_{||} & & &\\ \hline
 &  &-\lambda^\pm_{sc}-A&&\\
&\ldots &&-\lambda^\pm_{sc}-A &\\
 & &&&-\lambda^\pm_{sc}-A
\end{array}\right)
$$
respectively. As usual, we assume here that the spatial momentum is directed along
the first axis.

\subsection{The origin of DVZ discontinuity}

What does this mean?
The Newton potential, describing interaction between
$T_{00}$-components of the stress tensor can be read off from formula (\ref{TT})
and looks like
\be
U(r) \sim \frac{e^{-mr}}{r^{d-3}} -
\frac{1}{(d-1)(d-2)}\left( (d-2) \frac{e^{-mr}}{r^{d-3}} +
\frac{e^{-Mr}}{r^{d-3}}\right)
\ee
with
\be
m^2 = A\ \ \ \ {\rm and} \ \ \ \
M^2 = \frac{dAB-A^2}{(d-2)(A-B)}
\ee
Whenever
\be
A,B \rightarrow 0,\ \ {\rm potential} \ \
U(r)\rightarrow \frac{1}{r}\left(1-\frac{1}{d-2}\right)
=\frac{d-3}{d-2}\cdot\frac{1}{r^{d-3}},
\label{maslNewt}
\ee
which is also the massless-gravity value,
except for the special PF case when simultaneously
\be
\frac{A-B}{A^2} \rightarrow 0\ \ \ {\rm  and}
\ \ \ M\rightarrow \infty:
\ \ \
{\rm then}\ \ U(r) \rightarrow \frac{1}{r^{d-3}}
\left(1-\frac{1}{d-1}\right)
= \frac{d-2}{d-1}\cdot\frac{1}{r^{d-3}}
\label{PFNewt}
\ee
This is the DVZ discontinuity.

In other words, this is the
discontinuity due to the different limits
of the $a_5$ coefficient in the propagator,
\be
a_5 = -\frac{1}{k^2+A}\frac{k^2+\frac{AB}{A-B}}
{(d-2)k^2+\frac{dAB-A^2}{(A-B)}}
= -\frac{1}{(d-1)(d-2)}\left(\frac{d-2}{k^2+m^2}
+ \frac{1}{k^2+M^2}\right)=\\=
\left\{\begin{array}{ll}
-\frac{1}{(d-1)}\frac{1}{k^2}\hfill &\frac{A-B}{A^2} \to 0,\ \ A\to 0\\
-\frac{1}{(d-2)}\frac{1}{k^2}\ \ \ \ \ \ \ \ \ \ \ \ \ \ \ \ \ &A,B\to 0\\
\end{array}\right.
\label{a5}
\ee

In formal terms, the story is about the limit of a function
\be
\frac{ax+by}{cx+dy}
\ee
as $x,y\rightarrow 0$: the limit depends on the ratio
$x/y$. It is important that this ambiguity is intimately related
to the singularity of the function along the line $cx+dy=0$.

In physical terms, in PF case the ghost has infinite mass and
decouples, but in generic situation its mass tends
to zero along with graviton's mass, and they {\it both}
contribute to the Newton potential.
When masses exactly coincide (in particular, vanish)
ghost simply subtracts/adds to
the graviton, but in different way for different $a_k$-structures.
In particular it {\it adds} in $a_5$, which controls the Newton
potential.

\newpage

\section{Abandoning Lorentz invariance
\label{LV}}
\setcounter{equation}{0}

Lorentz violation as a phenomenologically viable
possibility was suggested by V.A.Kostelecky
and S.Samuel in \cite{KS} (see also \cite{LVm}) already long ago,
but it acquired enormous attention quite recently,
see \cite{LV}
for incomplete lists of papers
about its possible role in particle physics
and cosmology.
This paper is focused on
theoretical rather than
phenomenological aspects of Lorentz violation,
which were also addressed in \cite{LV}
and partly reviewed in \cite{RT}.

\subsection{Generalities}

Lorentz violation implies that the global symmetry
group $SO(1,d-1)$ is  broken down to $SO(d-1)$.
Surprisingly or not, this leads to rather drastic
changes in the structure of the theory,
revealing all the results of breakdown of gauge
invariance which remained hidden in Lorentz invariant
case.
In particular, since different reference frames are
no longer equivalent, the $SO(d-1)$ symmetry of
the spectrum in the rest frame is broken down to
helicity symmetry $SO(d-2)$ in all other frames,
what gives rise to highly non-trivial dispersion
relations for the elementary constituents (different
polarizations) of the gravity field.
These slightly unusual features remain obscured in
the case of massive vectors and come out of the shadow
only for tensor fields.

\bigskip

From $h_{\mu\nu}$ and $k_\mu$ one can make:
\begin{itemize}
\item
$4$ $h$-linear $SO(d-1)$-scalars:
$h_{00}$, $h_{ii}$, $k_ih_{0i}$,
$k_ik_jh_{ij}$,
\item
$2$ $h$-linear $SO(d-1)$-vectors:
$h_{0i}$, $k_ih_{ij}$  and
\item
$1$ $h$-linear $SO(d-1)$-tensor: $h_{ij}$
\end{itemize}
and thus $14 = \frac{4\cdot 5}{2} +
\frac{2\cdot 3}{2} + 1 = 10 + 3 + 1$ $h$-bilinear structures
which contribute into the propagator ${\cal P}_{\mu\nu,\alpha\beta}$.
The current-current interaction then looks like
\be
{\cal P}_{\mu\nu,\alpha\beta}T^{\mu\nu}T^{\alpha\beta}=\\
= a_{10}(k_iT_{0i})^2  + a_{11}(k_iT_{ij})^2 +
a_{20}T_{00}(k_ik_jT_{ij}) + a_{21}T_{ii}(k_ik_jT_{ij})
+ \\
+ a_{40}T_{0i}^2 + a_{41}T_{ij}^2
+ bT_{00}^2 + a_{50}T_{00}T_{ii} +
a_{51}(T_{ii})^2 +\\
+ c_1T_{00}(k_iT_{0i}) + c_2T_{ii}(k_jT_{0j})
+ c_3T_{0j}(k_iT_{ij}) +\\
+ a_{60}(k_iT_{0i})(k_ik_jT_{ij})+a_{61}(k_ik_jT_{ij})^2
\label{progra}
\ee
Our notation takes into account that each
structure with coefficients $a_k$ in the Lorentz invariant
case is now split into two and $a_{k0}$, $a_{k1}$ are the two
independent coefficients. The coefficient $b$ corresponds to what
was a combination of $a_4$ and $a_5$ and $c_i$ correspond to new emerged structures.
Note that $a_3$ is absent at all here, since we no longer keep
the parameter $\alpha$ (in the previous section we saw it
played no important role in massive cases) and define the propagator just by the
equation $K{\cal P}=E$. This would correspond to $a_2=a_3$.

Two of these structures, $[khk]^2$ and $[khk][kh]$
do not appear in the Lagrangian, because they are
more than quadratic in momenta. Other structures enter the Lagrangian with adjusted
coefficients so that the Lorentz invariance is not broken in
$k^2$-terms.
By now conventional parametrization of quadratic Lagrangian with
manifest Lorentz and general covariance violation in the mass
matrix is
$$K_{\mu\nu,\alpha\beta}h^{\mu\nu}h^{\alpha\beta} =$$
\centerline{$
=\left\{\Big( k_\mu k_\alpha\eta_{\beta\nu} +
k_\mu k_\beta\eta_{\alpha\nu} +
k_\nu k_\alpha\eta_{\beta\mu} +
k_\nu k_\beta\eta_{\alpha\mu}\Big) 
- k^2\Big(\eta_{\mu\alpha}\eta_{\nu\beta} +
\eta_{\nu\alpha}\eta_{\mu\beta}\Big)
- 2\Big(k_\mu k_\nu\eta_{\alpha\beta} +
k_\alpha k_\beta \eta_{\mu\nu}\Big)
+2k^2\eta_{\mu\nu}\eta_{\alpha\beta}\right\}h^{\mu\nu}h^{\alpha\beta}
+ $}
\be
+ 2m_0^2h_{00}^2 + 4m_1^2h_{0i}^2 - 2m_2^2h_{ij}^2
+ 2m_3^2 h_{ii}^2 - 4m_4^2h_{00}h_{ii}
\label{mact}
\ee
Lorentz invariance is restored provided
\be
m_0^2 = B-A,  \\
m_1^2=m_2^2=A,  \\
m_3^2=m_4^2=B
\ee

The propagator (\ref{progra}) is obtained by inverting the
square matrix in the action,

\bigskip

\begin{equation}
\left[\begin{array}{l|ccc|cc|ccc}
&00 & 01 & 11 & 0a & 1a & aa & ab & bb \\
\hline
00&m_0^2&&-m_4^2&&&-k_{||}^2-m_4^2&&-k_{||}^2-m_4^2\\
01&&m_1^2&&&&-\sqrt{2}\omega k_{||}&&-\sqrt{2}\omega k_{||}\\
11&-m_4^2&&m_3^2-m_2^2&&&-\omega^2+m_3^2&&-\omega^2+m_3^2\\
\hline
0a&&&&k_{||}^2+m_1^2&\omega k_{||}&&&\\
1a&&&&\omega k_{||}&\omega^2-m_2^2&&&\\
\hline
aa&-k_{||}^2-m_4^2&-\sqrt{2}\omega k_{||}&-\omega^2+m_3^2&&
&m_3^2-m_2^2&&-\omega^2+k_{||}^2+m_3^2\\
ab&&&&&&&\omega^2-k_{||}^2-m_2^2&\\
bb&-k_{||}^2-m_4^2&-\sqrt{2}\omega k_{||}&-\omega^2+m_3^2&&
&-\omega^2+k_{||}^2+m_3^2&&m_3^2-m_2^2\\
\hline
\end{array}\right]
\label{LVKM}
\end{equation}

\bigskip

or, in the rest frame,

\bigskip

\begin{equation}
\left[\begin{array}{l|cc|ccc}
&00 & 0i & ii & ij & jj \\
\hline
00&m_0^2&0&-m_4^2&0&-m_4^2\\
0i&0&m_1^2&0&0&0\\
\hline
ii&-m_4^2&0&m_3^2-m_2^2&0&-\omega^2+m_3^2\\
ij&0&0&0&\omega^2-m_2^2&0\\
jj&-m_4^2&0&-\omega^2+m_3^2&0&m_3^2-m_4^2\\
\hline
\end{array}\right]
\label{rfLVKM}
\end{equation}

\bigskip

For special values of masses the theory acquires some
residual gauge invariance \cite{Du}:
\be
\begin{array}{cc}
x^i \rightarrow x^i + \zeta^i(x^i,t):
& {\rm if}\ \ m_1=m_2=m_3=m_4=0 \\
t \rightarrow t + \zeta^0(x^i,t):
& {\rm if}\ \ m_0=m_1=m_4 = 0 \\
x^i \rightarrow x^i + \zeta^i(t):
& {\rm if}\ \  m_1=0
\end{array}
\ee
PF gravity corresponds to $m_0=0$, $m_1^2=m_2^2=m_3^2=m_4^2$.
In general the graviton mass is $m_2$ and the corresponding
$\frac{d(d-3)}{2}$ transverse modes are always split from
everything else, while the role of the other mass-parameters
is to control non-trivial dispersion relations for the other
constituents of the gravity field and their severe
(and generically inseparable) intermixing.
A lot of these complexities disappear if $m_0=0$ and such
model exhibits only minor deviations from the intuition
developed in experience with Lorentz invariant theories.

\subsection{Rubakov's approach}

In the pioneering paper \cite{Ru} V.Rubakov made the first
attempt to diagonalize the action (\ref{mact}).

He expressed $h_{\mu\nu}$ and the action through transverse
fields ($\chi_{ij}$, $u_{ij}$, $\psi$, $\tau$)
and additional Stueckelberg fields ($s_i$, $v$, $\sigma$)
and, next, through the gauge invariant vector
$w_i$ and scalars $\Phi$ and $\tau$:
\be
\begin{array}{ll}
h_{00} = \psi, &  \\
h_{0i} = u_i + k_i v, \ \ \ \ & k_iu_i = 0, \\
h_{ij} = \tau\delta_{ij}
+ \chi_{ij} + (k_is_j + k_js_i) + k_ik_j\sigma,
\ \ \ \ & k_i\chi_{ij} = k_j\chi_{ij} = 0,\ \ \ \chi{ii}=0, \ \ \ k_is_i=0
\end{array}
\ee
Since gauge transformations
$h_{\mu\nu} \rightarrow
h_{\mu\nu} + k_\mu\Xi_\nu + k_\nu\Xi_\mu$\ with\
$\Xi_0=\xi_0$,\ $\Xi_i = \xi_i + k_i\zeta$ and\ $k_i\xi_i=0$
act on these fields as
\be
\chi_{ij} \rightarrow k_i\xi_j+k_j\xi_i,  \\
u_i \rightarrow u_i + \omega\xi_i,  \\
s_i \rightarrow s_i + \xi_i,  \\
\psi \rightarrow \psi + 2\omega\xi_0,  \\
v \rightarrow v + \omega\zeta + \xi_0,  \\
\sigma \rightarrow \sigma + 2\zeta,  \\
\tau \rightarrow \tau
\ee
the gauge invariant vector and scalars are\footnote{In order to make a contact
of these notations from \cite{Ru} with notations of the paper \cite{RT}:
$$
\begin{array}{ccccccc}
\tau=2\psi&
\chi_{ij}=h^{TT}_{ij}&
s_i=-F_i&
\sigma=2E&
u_i=S_i&
v=B&
\psi=2\phi
\end{array}
$$
}
\be
w_i = u_i - \omega s_i,  \\
\Phi = \psi - 2\omega v + \omega^2\sigma,  \\
\tau
\ee
and the action {\it for conserved stress-tensor} $T_{\mu\nu}$,
$k^\mu T_{\mu\nu}=0$ is
\be
-k^2\chi_{ij}^2 -(d-1)(d-2)\omega^2\tau^2 + 2\vec k^2 (w_i^2 -(d-2)\Phi\tau +
{(d-2)(d-3)\over 2}\tau^2)
\ \ \ \ \ \ 
+\ \ \chi_{ij}T_{ij} - 2w_iT_{0i} + \Phi T_{00} + \tau T_{ii} +  \\
+ m_0^2\psi^2 + 2m_1^2(u_i^2 + \vec k^2 v^2) 
- m_2^2\Big(\chi_{ij}^2 + (d-1)\tau^2 + 2\vec k^2(s_i^2+\sigma\tau)
+ \vec k^4 \sigma^2 \Big)
+ m_3^2(\vec k^2\sigma + (d-1)\tau)^2 - 2m_4^2\psi (\vec k^2 \sigma + (d-1)\tau)
\ee
The first line is quadratic part of Einstein-Hilbert action
and it contains nothing but gauge invariant fields.
In general relativity the scalar $\Phi$ does not have a kinetic term
and does not propagate.
The second line contains gauge and Lorentz-violating terms
and depends on all the fields, in particular, provides
kinetic terms for Stueckelberg fields $s_i$, $v_i$ and $\sigma$
(if $m_0^2$ and $m_1^2$ are both non-vanishing).

It is rather straightforward to diagonalize this Lagrangian
and analyze its particular eigenvectors: modes of the gravity
field. This was done in \cite{Ru} under a strongly simplifying
assumption $m_0=0$ (which also guarantees the absence of ghosts),
the same kind of analysis in general situation being rather
tedious, and only the Stueckelberg sector was studied in \cite{Du} for
$m_0\ne 0$.
In what follows we use a slightly different technique,
which also has an advantage of being easily made algorithmic and
thus allows one to make tedious calculations in a systematic way
with the help of a computer. In fact, as in the previous section,
we use two different ways of analyzes: both though manifest constructing
the propagator and through immediate diagonalizing the kinetic matrix.

\subsection{Particle content of the theory \label{degref}}

Gravitational field is described by symmetric $d\times d$ matrix

As in the Lorentz invariant case, the  $\frac{d(d+1)}{2}$
components of the gravitational fields split into
$\frac{(d+1)(d-2)}{2}$ components describing
a gauge-invariant traceless graviton, one scalar trace and
$d$ components of gauging (Stueckelberg) vectors, which is in turn
split into space-time transverse vector and the secondary Stueckelberg
scalar, see eq.(\ref{degf0}) above.
In the rest frame (which exists since the fields are massive)
these $\frac{(d+1)(d-2)}{2}$ components are all degenerate,
being related by the action of $SO(d-1)$ symmetry group.
However, if Lorentz invariance is broken, degeneration is
lifted for $\vec k\neq 0$ and the
$\frac{(d+1)(d-2)}{2}$ components further split into
transverse graviton, transverse vector and a scalar,
which are representations of
the "helicity" group $SO(d-2)$
acting in the hyperplane, orthogonal to $\vec k$.

These degeneration properties are most concisely reflected in
determinant formula for the kinetic matrix
\be
\det K = D_{gr}^{\frac{d(d-3)}{2}}
D_{vec}^{d-2} D \ \stackrel{\vec k = 0}{\longrightarrow} \
m_1^{2(d-2)}D_{gr}^{\frac{(d+1)(d-2)}{2}} D_{tr}
\label{detk}
\ee
where
\be
D_{vec} = -m_1^2k_0^2+m_2^2\vec k^2 + m_1^2m_2^2 =
m_1^2\left( -k_0^2+\frac{m_2^2}{m_1^2}\vec k^2 + m_2^2\right),
\label{D1form}
\ee
\be
D_{gr} = -k_0^2 + \vec k^2 + m_2^2
\label{D2form}
\ee
and $D$ is a sophisticated expression of power $4$ in $k_0$ and $|\vec k|$:
\be
D= (d-2)m_0^2m_1^2k_0^4
+\Big((2d-4)m_0^2m_3^2 -(2d-4)m_0^2m_2^2
+(2d-4)m_1^2m_4^2 -(2d-4)m_4^4\Big)k_0^2\vec k^2 +  \\
+\Big((d-2)m_1^2m_3^2 -(d-2)m_1^2m_2^2\Big)\vec k^4 + \\
+\Big((d-1)m_4^4m_1^2-(d-3)m_2^2m_0^2m_1^2 -(d-1)m_3^2m_0^2m_1^2\Big)k_0^2
+  \\
+\Big((d-3)m_0^2m_2^2m_1^2 -(d-3)m_3^2m_0^2m_1^2 -(2d-4)m_4^2m_2^2m_1^2
+(d-3)m_4^4m_1^2\Big)\vec k^2+\\
+(d-1)m_2^2m_0^2m_3^2m_1^2 -m_2^4m_1^2m_0^2 -(d-1)m_2^2m_1^2m_4^4
\label{Dform}
\ee
For $d=4$ (\ref{Dform}) gives
\be
D= 2m_0^2m_1^2k_0^4
+(-4m_0^2m_2^2 +4m_0^2m_3^2 +4m_1^2m_4^2 -4m_4^4)k_0^2\vec k^2
+(2m_1^2m_3^2 -2m_1^2m_2^2)\vec k^4 + \\
+(-m_2^2m_0^2m_1^2 -3m_3^2m_0^2m_1^2 +3m_4^4m_1^2)k_0^2
+(m_0^2m_2^2m_1^2 -m_3^2m_0^2m_1^2 -4m_4^2m_2^2m_1^2 +m_4^4m_1^2)\vec k^2+\\
+3m_2^2m_0^2m_3^2m_1^2 -m_2^4m_1^2m_0^2 -3m_2^2m_1^2m_4^4
\label{Dform4}
\ee
At $\vec k=0$ our $D_{vec}$ turns into $D_{gr}$ and $D$ decomposes into two factors,
one of which is also $D_2$:
\be
D_{vec} \ \stackrel{\vec k = 0}{\longrightarrow} \ m_1^2 D_{gr},  \\
D \ \stackrel{\vec k = 0}{\longrightarrow} \ D_{gr}D_{tr},  \\
D_{tr} = m_1^2\left(m_0^2m_2^2+(d-1)m_4^4+(d-2)m_0^2\omega^2-(d-1)m_0^2m_3^2\right)
\ee
Clearly a drastic simplification occurs also in the case of $m_0=0$ \cite{Ru},
when $D$ turns into
\be
D_{m_0=0} = 4m_4^2(m_1^2 -m_4^2)k_0^2\vec k^2
+2m_1^2(m_3^2 -m_2^2)\vec k^4 +3m_1^2m_4^4k_0^2
+m_1^2m_4^2(m_4^2-4m_2^2)\vec k^2 -3m_1^2m_2^2m_4^4
\ee

One more implication of (\ref{detk}) is that only degrees of freedom
from the first braces in expansion (\ref{degf0}) are dynamical fields,
even after violation of the Lorentz and gauge symmetries.

\subsection{Normal modes for gravity fields}

We manifestly describe the normal modes only in the rest frame in this case,
since, formulas in the moving frames
become very involved and non transparent. Note that
Lorentz transformation can no longer be used to deduce formulas in the moving
frame. Therefore, we explicitly present and discuss, at least, eigenvalues
in the generic frame. Note that the rest frame analysis is already enough
to see ghosts.

Rest frame, where $k_{||}=0$ is distinguished because $SO(d-1)$ symmetry
is fully restored in it and there is no reason to distinguish between
$1$ and other spatial directions.

Its {\bf eigenvectors} ($d\times d$ matrices) {\bf and eigenvalues} are:
\begin{equation}
\begin{array}{ccc}
\omega^2 -m_2^2 &\left(\begin{array}{cc|c}
0 & 0 & 0 \\
0 & 0 & 0 \\ \hline
0 &  0 & h_{ab}
\end{array}\right) & {\rm transverse\ traceless\ graviton} \\
\\
\omega^2 -m_2^2&\left(\begin{array}{cc|ccc}
0 & 0 & \dots&0&\ldots \\
0 & 0 & \ldots&1&\ldots \\ \hline
&\ldots\\
0 &  1 & &0\\
&\ldots
\end{array}\right) & {\rm transverse}\ (d-2)-{\rm vector} \\
\\
m_1^2&\left(\begin{array}{cc|ccc}
0 & 0 & \ldots&1&\ldots \\
0 & 0 & \ldots&0&\ldots\\ \hline
&\ldots\\
1 &  0 && 0\\
&\ldots
\end{array}\right) & {\rm Stueckelberg}\ (d-2)-{\rm vector} \\
\\
m_1^2&
\left(\begin{array}{cc|c}
0& 1  & 0 \\
1 & 0  & 0 \\ \hline
0 &  0 & 0
\end{array}\right)&
 {\rm Stueckelberg\ scalar}\\
\\
\omega^2-m_2^2&\left(\begin{array}{cc|c}
0 & 0 & 0 \\
0 & d-2 & 0 \\ \hline
0 &  0 & -1
\end{array}\right) & {\rm secondary\ Stueckelberg\ scalar}\\
\\
\lambda_{sc}^{\pm}&\left(\begin{array}{cc|c}
-(d-1)m_4^2 & 0 & 0 \\
0 & \lambda_{sc}^{\pm}-m_0^2  & 0 \\ \hline
0 &  0 & \lambda_{sc}^\pm-m_0^2
\end{array}\right) & {\rm two\ mixed\ traces,\ spatial\ and\ space-time} \\
\end{array}
\end{equation}
or, if we do not distinguish between the $1$ and other space directions,
\be
\begin{array}{ccc}
\omega^2-m_2^2&\left(\begin{array}{c|c}
0 & 0 \\ \hline
0 & h_{ij}
\end{array}\right) & {\rm  traceless\ graviton} \\
m_1^2&\left(\begin{array}{c|c}
0 & 1 \\ \hline
1 &  0
\end{array}\right) & {\rm Stueckelberg}\ (d-1)-{\rm vector} \\
\lambda_{sc}^\pm&\left(\begin{array}{c|c}
-(d-1)m_4^2 & 0 \\ \hline
0& \lambda_{sc}^\pm-m_0^2
\end{array}\right) & {\rm
a\ mixture\ of\ secondary\ Stueckelberg\ scalar\
and\ space-time\ trace} \\
\end{array}
\ee
In other words, in the massive gravity one has the three possible kinds
of the eigenvalues in the rest frame:
\be
\begin{array}{ccc}
&\lambda_{gr} = \omega^2- m_2^2 & \frac{(d+1)(d-2)}{2}\ {\rm times} \\
& \lambda_{vec} = m_1^2 & d-1\ {\rm times} \\
& \lambda_{sc}^\pm = \xi+m_0^2\pm\sqrt{\xi^2+(d-1)m_4^4}
& {\rm two\ traces}\\
&\xi\equiv
{(d-1)m_3^2\over 2}-{m_0^2+m_2^2\over 2}-{(d-2)\omega^2\over 2}
\end{array}
\ee
The last two eigenvalues, $\lambda_{sc}^\pm$ describe only one propagating
mass. Indeed, the zeroes
of the {\it two} equations $\lambda_{sc}^\pm=0$ are encoded in the equation
$\lambda^+_{sc}\lambda^-_{sc}=2\xi m_0^2+m_0^4-(d-1)m_4^4=0$, which is linear
in $\omega^2$. This mass is manifestly given by
\be\label{M2}
M^2={d-1\over d-2}\left(m_3^2-{m_4^4\over m_0^2}\right)-{m_2^2\over d-2}
\ee
Therefore, in this case of broken Lorentz symmetry, similarly to the Lorentz invariant
case, there are two propagating masses, $M^2$ and $m^2=m_2^2$, although the
dispersion laws are far more tricky in the present case, as we shall see in s.5.6
(e.g. some of these modes propagate with another speed of light). Note that the two
masses coincide, $M^2=m^2$ provided
\be
(m_3^2-m_2^2)m_0^2=m_4^2
\ee
Now, similarly to the Lorentz invariant case, one immediately observes a ghost:
since the derivative $\left|{\partial\lambda(\omega^2)\over\omega^2}\right|_{\lambda^\pm_{sc}}$
is negative at that single point where $\lambda(\omega^2)$ crosses the abscissa axis.
This ghost may not propagate if the parameters are chosen so that $M^2<0$. However, then
one has to look at non-zero spatial momentum to see if this crossing point
$\lambda^\pm_{sc}$ becomes positive.

As before, one
may try to remove this ghost from the spectrum bringing its mass to infinity,
which is equivalent to putting $m_0=0$ provided the both $m_0$ and $m_4$ are finite,
or either $m_0$ or $m_4$ go to infinity.
Therefore, $m_0=0$ is, at least, a sufficient condition for ghost free space-time, \cite{Ru}!

\subsection{Characteristic polynomials}

Now let us analyze the modes in general frame looking at the eigenvalues.
Here we present the analysis in terms of
the Euclidean normal modes (using the Lorentz normal modes this time does not lead to
any simplifications, since the Lorentz symmetry is violated anyway).

One of the eigenvalues corresponds to the graviton propagator and is
\be
D_{gr} = -\lambda_{gr} = -\omega^2+\vec k^2 + m_2^2 =  k^2 + m_2^2,
\ee
It describes the ${d(d-3)\over 2}$ graviton modes.

The Stueckelberg and transverse vectors are described now
by the $2\times (d-2)$ eigenvalues\footnote{One can compare these formulas with
those for the Lorentz normal modes,
\be
\lambda^\pm_{vec}=-{k^2+m_1^2+m_2^2\pm\sqrt{(m_2^2-m_1^2)^2
+2(m_1^2-m_2^2)(\omega^2+\vec k^2)+k^4}\over 2}
\ee
One can see there are no any simplifications in this case.
}
\be
\lambda_{vec}^\pm = \frac{1}{2}\left(\omega^2+\vec k^2 + m_1^2-m_2^2
\pm \sqrt{(\omega^2+\vec k^2)^2 + 2(-\omega^2+\vec k^2)(m_1^2+m_2^2)
+ (m_1^2+m_2^2)^2}
\right) =  p_2 \pm \sqrt{p_4},
\ee
which combine into the polynomial
\be
D_{vec} = -\lambda^+_{vec}\lambda^-_{vec} = p_4-p_2^2
= -m_1^2\omega^2 + m_2^2\vec k^2 + m_1^2m_2^2
\ee
and are described by a characteristic polynomial
\be
P_2(\lambda) = \lambda^2 - 2p_2\lambda - D_{vec}
\ee
The remaining four scalars combines into quite involved quartic
characteristic polynomial of the form:
\be
Q_4(\lambda) = \lambda^4+
\lambda^3\Big((d-3)(\omega^2-\vec k^2-m_3^2)- m_0^2-m_1^2 +2m_2^2- 2m_3^2\Big)
 +  \\
+\lambda^2 \Big\{-(d-2)(\omega^2+\vec k^2)^2
+\omega^2\Big(2m_3^2+(d-3)(-m_0^2-m_1^2+m_2^2+m_3^2)\Big)+\\
+\vec k^2\Big(-2m_4^2+(d-3)(m_0^2+m_1^2-m_2^2+m_3^2-2m_4^2)\Big) +
\\
+\Big((m_0^2m_1^2+m_2^4-2m_0^2m_2^2-2m_1^2m_2^2)
+(d-1)(-m_4^4-m_2^2m_3^2+m_1^2m_3^2+m_3^2m_0^2)\Big)\Big\}
\\
+\lambda\Big\{(d-2)\Big(\omega^4(m_0^2+m_1^2)
+2\omega^2\vec k^2(m_0^2-m_2^2+m_3^2+m_4^2)
+\vec k^4(m_1^2-m_2^2+m_3^2)\Big) +
\\
+\omega^2\Big((d-1)(m_4^4-(m_0^2+m_1^2)m_3^2)+
       (d-3)(-m_1^2m_2^2+m_0^2m_1^2-m_0^2m_2^2)\Big) +
      \\
+\vec k^2\Big(2(d-2)(m_1^2m_4^2-m_2^2m_4^2)+
       (d-3)(m_0^2m_2^2+m_1^2m_2^2-m_0^2m_1^2-m_1^2m_3^2-m_3^2m_0^2+m_4^4)\Big)
       + \\
+\Big(2m_0^2m_1^2m_2^2-m_0^2m_2^4-m_1^2m_2^4+
(d-1)(m_0^2m_2^2m_3^2+m_4^4m_1^2-m_2^2m_4^4
-m_0^2m_1^2m_3^2+m_1^2m_2^2m_3^2)\Big)\Big\} +
\\
+\Big\{(d-2)\Big(-\omega^4m_0^2m_1^2
+2\omega^2\vec k^2(m_4^4-m_3^2m_0^2+m_0^2m_2^2-m_1^2m_4^2)
+\vec k^4(m_1^2m_2^2-m_1^2m_3^2)\Big) +\\
+\omega^2\Big((d-1)(m_0^2m_1^2m_3^2-m_4^4m_1^2)
+(d-3)m_0^2m_1^2m_2^2\Big) + \\
+\vec k^2\Big(2(d-2)m_1^2m_2^2m_4^2
+(d-3)(-m_0^2m_1^2m_2^2+m_0^2m_1^2m_3^2-m_4^4m_1^2)\Big)+ \\
+\Big(m_2^4m_0^2m_1^2
+(d-1)(m_2^2m_4^4m_1^2-m_2^2m_3^2m_0^2m_1^2)\Big)\Big\}\label{Q4}
\ee
so that $Q_4(0)=-D$. This quantity is definitely the same in the Lorentz and
Euclidean modes, since this comes from the determinant of the kinetic operator
in the scalar mode sector (similarly, are invariant $D_{gr}$ and $D_{vec}$).

In the rest frame, where $\vec k = 0$, we definitely returns to formulas of the
previous subsection:
\be
\lambda_{gr} = \omega^2-m_2^2,  \\
\lambda_{vec}^+ = \omega^2 - m_2^2\\
\lambda_{vec}^- = m_1^2
\ee
and $Q_4(\lambda)$ factorizes, in accordance with (\ref{rfqq}):
\be
Q_4(\lambda) =
(\lambda-m_1^2)\Big(\lambda-(\omega^2-m_2^2)\Big)\cdot  \\
\cdot
\Big\{\lambda^2-\lambda\Big(m_0^2-m_2^2+(d-1)m_3^2 - (d-2)\omega^2\Big)
-(d-2)\omega^2 m_0^2-m_0^2m_2^2-(d-1)m_4^4+(d-1)m_0^2m_3^2\Big\}
\ee

Therefore, one can see in the moving frame that, of the ${(d+1)(d-2)\over 2}$
modes in the rest frame with $\lambda=\omega^2-m_2^2$, ${d(d-3)\over 2}$
remains graviton modes, while $d-2$ modes along with $d-2$ Stueckelberg modes
(with $\lambda=m_1^2$) compose the $D_{vec}$, and one mode with
$\lambda=\omega^2-m_2^2$, one mode with $\lambda=m_1^2$ and two trace modes
altogether compose $D$ (or $Q_4(\lambda)$).

\subsection{Ghosts, tachyons and others}

With the characteristic polynomial in hands, we can discuss the
properties -- and peculiarities -- of the spectrum of
Lorentz-violating gravity. Basically, there are three different important peculiarities:
when there is a ghost, when there is a tachyon and when the speed of propagation
of a mode differs from the light speed. First of all, the tensor sector is healthy
of all these phenomena.

\paragraph{Superluminal propagation.} One can most directly
observe the superluminals in the vector
sector of the theory. Indeed, the dispersion law $D_{vec}=0$ implies that the
vectors propagate with the speed ${m_2\over m_1}$. This speed can be larger or less
than the speed of graviton (which is 1) depending on the relation between masses.

However, the main peculiarities are related to the scalar sector of the theory.

\paragraph{Ghost.} We already discussed the appearance of a ghost
in the scalar sector in s.5.4 and
concluded that it may have infinite mass (i.e. disappears from the spectrum
of the {\it linearized} gravity, but can easily come back beyond quadratic
approximation as a
Boulware-Deser mode), provided $m_0^2=0$. Otherwise, there may be a ghost, at least,
at $\vec k=0$ and $M^2\ge 0$.
In other words, there will be a mode which is constant in space and would grow in time
(simultaneously in the whole space). Still, this is not a ghost propagating in space.

In order to move slightly away from the $\vec k^2=0$ point, i.e. consider propagation in space,
one can find a solution to (\ref{Q4}) at small $\vec k^2$. The limit of $\vec k^2$ is smooth,
and one would come to a ghost again unless $m_1=0$. Indeed, one expects a singularity in this
latter case, since the determinant (\ref{detk}) is proportional to $m_1$ at zero momentum.

Let us make a closer inspection of this particular case. In this case the dispersion law
looks quite strange
\be
D\sim\omega^2\vec k^2=0
\ee
This corresponds to infinite speed of light for
the vector mode and leads to a very exotic
excitation, being a carrier of instantaneous
interaction $\omega^2=0$.
One can actually describe the slightly-virtual (small values of $\omega^2$)
\textit{instantaneon} analytically: the corresponding eigenvalue is \be\label{m10}
\lambda_{instant} = \frac{2(d-2)\Delta \cdot \omega^2\vec
k^2}{(d-2)(m_3^2-m_2^2)\vec
k^4-\left[(d-3)\Delta+2(d-2)m_2^2m_4^2\right]\vec
k^2+(d-1)m_2^2(m_0^2m_3^2-m_4^4)-m_0^2m_2^4}+ O(\omega^4) \ee
where
\be\label{Delta}
\Delta \equiv m_0^2(m_3^2-m_2^2) - m_4^4 = 0
\ee
This excitation is simultaneously a ghost, $d\lambda/d\omega^2<0$ on
shell, when $\vec k^2$ lies in between the zeroes of the denominator
in (\ref{m10}).

Thus, in the specific case of $m_1=0$ the ghost is
related with excitations with the specific dispersion law, and is not a particle-like
ghost. This is probably the reason why it was not recognized as a ghost in \cite{Du}.
It would be interesting to better understand physical implications of this instantaneon.

\paragraph{Tachyons.} Now we return to our discussions of the ghost-free regime at $m_0=0$.
In order to see if there are tachyons, one has to put
$\omega=0$ and look for real solutions $\lambda(\vec k)=0$. Note that this is not
the same as to look for a mode with negative mass square because of the tricky
dispersion law. Indeed, the mass describes the pole of the propagator in
$\omega$ at zero $\vec k$, while the tachyon has to do with its $\vec k$-dependence.

In order to guarantee
the absence of tachyons in the vector and tensor sectors, one has to require
$m_1^2\ge 0$ and $m_2^2\ge 0$. Then, it is again enough
to look at the product of all 4 scalar eigenvalues, which is $D$. Thus, the tachyon
is absent as soon as there is no real-valued solution of the equation
\be\label{tach}
-\left.D(\omega^2,\vec k^2)\right|_{\omega=0}=\rho \vec k^4+\eta\vec k^2+\zeta=0
\ee
where
\be
\rho\equiv (d-2)(m_1^2m_2^2-m_1^2m_3^2)\\
\eta\equiv 2(d-2)m_1^2m_2^2m_4^2
+(d-3)(-m_0^2m_1^2m_2^2+m_0^2m_1^2m_3^2-m_4^4m_1^2)\\
\zeta\equiv m_2^4m_0^2m_1^2
+(d-1)(m_2^2m_4^4m_1^2-m_2^2m_3^2m_0^2m_1^2)=-(d-2)m_0^2m_1^2m_2^2M^2
\ee
The tachyon is absent either if the discriminant of (\ref{tach}) is negative,
\be
\eta^2-4\rho\zeta<0
\ee
or if both solutions ${1\over 2}\left(-\eta\pm\sqrt{\eta^2-4\rho\eta}\right)$
of the quadratic equation,
(\ref{tach}) are negative
\be
\eta>0,\ \ \ \ \ \ \ \rho\zeta>0
\ee
If neither of these conditions is satisfied, there is a tachyon in the spectrum.
In order to have a theory both without the ghost and the tachyon, one can
put $m_0^2=0$ and require that
\be
m_2^2>m_3^2,\ \ \ \ \ \ \ 2(d-2)m_2^2>(d-3)m_4^2,\ \ \ \ \ \ m_4^2\ge 0
\ee
(these are the conditions obtained in \cite{Ru} in $d=4$) or
\be
\left[2(d-2)m_2^2-(d-3)m_4^2\right]^2<4(d-1)(d-2)m_2^2(m_2^2-m_3^2)
\ee
We assumed here that $m_1^2$, $m_2^2$ and $m_4^2$ are non-zero.

One can also consider the border cases. If one of the masses
$m_1^2$, $m_2^2$ or $m_4^2$ is zero, $\zeta=0$ which means that there is
a massless mode in the spectrum. Now one has to differ between different cases.

If $m_2^2=0$, there is a tachyon unless also $m_3^2=0$. In this latter case,
the speed of light of the vector mode becomes zero, and the dispersion law
\be\label{k4}
D\sim\vec k^4=0
\ee
implies the mode does not propagate in time at all.

There is also a possibility of $m_4^2=0$ that leads to a non-propagating mode as well,
with the same dispersion law (\ref{k4}).

The last border case to consider is $\rho=0$, i.e. $m_2^2=m_3^2$.
Then, the tachyon is absent if $\eta\ge 0$. This means
\be
2(d-2)m_2^2\ge (d-3)m_4^2, \ \ \ \ \ \ m_4^2\ge 0
\ee
In particular, if the equality is realized
in these formulas, $\eta=0$ and
\be
2(d-2)m_2^2= (d-3)m_4^2
\ee
the dispersion law
acquires the form
\be
2(d-2)(m_4^2-m_1^2)\omega^2\vec k^2-(d-1)m_1^2m_4^2\omega^2+
(d-1)m_1^2m_2^2m_4^2=0
\ee

\subsection{Dispersion relations}

One of spectacular puzzles of massive gravity is emergency
of non-trivial dispersion relations $\omega = \epsilon(|\vec k|)$.
This looks puzzling because usually does not happen
in the theory of some $N$ scalar massless fields perturbed
by an arbitrary mass matrix:
\be
\sum_{a=1}^N \Big(k^2\phi_a^2 + J_a\phi_a\Big)
+ \sum_{a,b=1}^N M_{ab}\phi_a\phi_b
\ee
In general kinetic and mass matrices
define two quadratic forms which {\it can} be
simultaneously diagonalized, but diagonalization of $M_{ab}$
breaks down diagonal form of the field-current coupling.
This phenomenon is well
known as the Kobayashi-Maskawa mixing in the Standard Model of
elementary particles \cite{LBO}.
Characteristic equation defining the eigenvalues of such quadratic
Lagrangian without currents,
\be
D_M = \det_{N\times N} \Big( k^2 \delta_{ab} + M_{ab} \Big) = 0
\ee
is actually a product
\be
D_M = \prod_{a=1}^N \Big(k^2 + m_a^2(M)\Big) =
\prod_{a=1}^N \Big(-\omega^2 + \epsilon_a^2(|\vec k|)\Big)
\label{prod}
\ee
so that dispersion law is the standard relativistic
$\omega = \epsilon_a(|\vec k|) = \sqrt{\vec k^2 + m_a^2}$.

If Lorentz symmetry $SO(d-1,1)$ is broken down to $SO(d-1)$,
there are in general {\it three} different matrices:
\be
-\sum_{a=1}^N \omega^2\phi_a^2 +
\sum_{a,b=1}^N \Big(N_{ab}\vec k^2 + M_{ab}\Big)\phi_a\phi_b
\ee
and characteristic equation, defining $\omega(|\vec k|)$, is
more sophisticated:
\be
D_{N,M} = \det_{N\times N} \Big( -\omega^2 \delta_{ab} +
N_{ab}\vec k^2 + M_{ab} \Big) = 0
\label{DMN}
\ee
While $D_{N,M}$ is still a product like the last formula
in (\ref{prod}), the roots
$\epsilon_a(|\vec k|)$ can now be highly non-trivial
functions of the space momentum $\vec k$.
This would explain the origin of non-trivial dispersion relations,
but the problem is that in massive gravity one does {\it not}
introduce any non-trivial matrix $N_{ab}\neq \delta_{ab}$:
all Lorentz violation is concentrated in massive matrix
and does not affect the kinetic term!
This seems to imply that nothing more than a KM mixing can
occur with no severe change to dispersion relations --
but this is actually {\it not} the case, as we see in
eqs.(\ref{D1form})-(\ref{Dform}) and in s.5.6.

Resolution of the puzzle is in the concept of
Stueckelberg fields: when mass matrix breaks some gauge
symmetry, it gives rise to {\it kinetic} terms for
the newly revived gauge degrees of freedom.
From the point of view of above scalar theory this looks strange:
if massive matrix involves more fields than the kinetic matrix,
\be
\sum_{a=1}^N \Big(k^2\phi_a^2 + J_a\phi_a\Big)
+ \sum_{a,b=1}^{N+n} M_{ab}\phi_a\phi_b
\ee
then the extra $n$ fields $\phi_a$ should be represented as
{\it derivatives},
$\phi_{a'} = \sum_{b'=1}^n k_\mu C^\mu_{a'b'}\varphi_{b'}$.
Then the new kinetic term involves new Stueckelberg fields
$\varphi_{a'}$, breaks Lorentz invariance and we come back to
the situation described in (\ref{DMN}), where non-trivial
dispersion relations are of no surprise.
The problem is that above substitution $\phi \rightarrow \varphi$
looks {\it ad hoc} and it is actually justified only
when $\phi_{a'}$ describe pure gauge degrees of freedom.

\subsection{Manifest expression for the propagator}

Now we come the current-current interaction. To this end, we need to know the
propagator. Coefficients in the propagator in 4 dimensions are:
\be
a_{10}
= \frac{2}{D_{vec}D}
\Big(( 4m_4^4 +2m_2^2m_0^2 -4m_0^2m_3^2  ) \omega^4
-4m_2^2m_4^2 \omega^2\vec k^2
+(2m_2^4 -2m_2^2m_3^2   )\vec k^4 +  \\
+(7m_2^2m_0^2m_3^2 -7m_2^2m_4^4 -3m_0^2m_2^4   ) \omega^2
+ (4m_2^4m_4^2 -m_2^4m_0^2 -m_2^2m_4^4 +m_3^2m_0^2m_2^2 ) \vec k^2
+ (3m_2^4m_4^4 -3m_0^2m_3^2m_2^4 +m_2^6m_0^2   )\Big),  \\
a_{11} = \frac{-2(-\omega^2+\vec k^2+m_1^2)}
{m_1^2(-\omega^2+\vec k^2+m_2^2)(-\omega^2+\frac{m_2^2}{m_1^2}\vec k^2 + m_2^2)}
= -2\frac{-\omega^2+\vec k^2+m_1^2}{D_{vec}D_{gr}},  \\
a_{20} 
=-2\frac{(2(m_1^2-2m_4^2)\omega^2+m_1^2m_4^2+m_1^2m_2^2-3m_1^2m_3^2)}
{D},
\\
a_{21} 
= \frac{2}{D_{gr}D}\Big(
( m_4^2m_1^2 +m_0^2m_1^2 +2m_3^2m_0^2 -2m_2^2m_0^2 -2m_4^4)\omega^2
+ (m_4^2m_1^2 +m_3^2m_1^2 -m_1^2m_2^2 )\vec k^2 +  \\
+ ( 2m_1^2m_4^4 + m_2^2m_1^2m_0^2 -m_4^2m_1^2m_2^2
-2m_3^2m_0^2m_1^2 )\Big), \\
a_{60} = \frac{-4\omega }{D_{vec}D}\Big(
(2m_2^2m_0^2 -4m_3^2m_0^2 -2m_4^2m_1^2 +4m_4^4)\omega^2
+ ( 2m_1^2m_2^2 -2m_3^2m_1^2 -2m_2^2m_4^2 )\vec k^2 +  \\
+(-m_2^2m_1^2m_0^2+m_3^2m_0^2m_1^2-m_2^4m_0^2+3m_2^2m_0^2m_3^2
+2m_4^2m_1^2m_2^2 -m_1^2m_4^4 -3m_2^2m_4^4)\Big), \\
a_{61} = \frac{1}{D_{vec}D_{gr}D}\Big(
( 4m_2^2m_0^2 +2m_1^4 +8m_4^4 -8m_1^2m_4^2 -8m_0^2m_3^2)\omega^4 +\\
+ (8m_1^2m_4^2 +8m_3^2m_0^2 -8m_4^4 -4m_0^2m_2^2 - 2m_1^4
-4m_3^2m_1^2+2m_1^2m_2^2)\omega^2\vec k^2 
+ (4m_1^2m_3^2 -2m_1^2m_2^2   ) \vec k^4 +\\
\ \ \ \ \ \ \ \ \
+ (  8m_1^2m_2^2m_4^2 -4m_2^2m_0^2m_1^2 -4m_4^4m_1^2 +4m_3^2m_0^2m_1^2
+ \\
+6m_2^2m_0^2m_3^2 -2m_0^2m_2^4 -6m_2^2m_4^4 -m_2^2m_1^4 +2m_1^4m_4^2
-3m_1^4m_3^2+m_1^4m_0^2) \omega^2 + \\
+ ( 2m_1^2m_4^4 +3m_2^2m_3^2m_1^2 +m_0^2m_2^2m_1^2
-2m_3^2m_0^2m_1^2-2m_4^2m_2^2m_1^2 -m_2^4m_1^2 + 4m_1^4m_3^2
-2m_2^2m_1^4 ) \vec k^2 + \\
+ ( 2m_1^4m_4^4 -2m_2^2m_1^4m_4^2  +m_2^2m_1^4m_0^2 +3m_2^2m_1^4m_3^2
-2m_1^4m_3^2m_0^2-m_2^4m_1^4)
\Big),  \\
b = -\frac{
(   -2m_1^2 )\omega^4 +
4(m_2^2-m_3^2) \omega^2\vec k^2+
(m_2^2m_1^2 +3m_3^2m_1^2)\omega^2
+ ( m_3^2m_1^2 -m_1^2m_2^2 )\vec k^2
+ ( m_2^4m_1^2 -3m_3^2m_2^2m_1^2)
}{D},  \\
a_{40} = 2\frac{-\omega^2+m_2^2}{D_{vec}},  \\
a_{41} = -\frac{1}{D_{gr}},
\ee
\be
a_{51} = \frac{-1}{D_{gr}D}\Big(
( m_0^2m_1^2  )\omega^4
+(  2m_4^2m_1^2 +2m_3^2m_0^2-2m_4^4 -2m_0^2m_2^2 ) \omega^2\vec k^2
+ m_1^2(m_3^2 -m_2^2)\vec k^4 + \\
+( m_4^4m_1^2 -m_2^2m_0^2m_1^2 -m_3^2m_0^2m_1^2 )\omega^2
+ (m_0^2m_2^2m_1^2 -2m_4^2m_2^2m_1^2 +m_4^4m_1^2 -m_3^2m_0^2m_1^2)\vec k^2
+ (m_2^2m_0^2m_3^2m_1^2 -m_2^2m_1^2m_4^4   )\Big),  \\
a_{50} = \frac{2m_1^2
\Big(-m_4^2\omega^2 -(m_3^2-m_2^2)\vec k^2 + m_2^2m_4^2\Big)}{D}, \\
c_1 = \frac{-8\omega(m_4^2\omega^2+(m_3^2-m_2^2)\vec k^2 -m_2^2m_4^2)}{D},  \\
c_2 = 4\omega\frac{(m_4^4+m_2^2m_0^2-m_3^2m_0^2)}{D},  \\
c_3 = \frac{4\omega}{D_{vec}}
\ee
Restoring the $d$-dependence, one can see that only a few coefficients slightly
depend on the space-time dimension:
\be
\hspace{-1.5cm}
{a_{10}}
= \frac{2}{D_{vec}D}
\Big\{\Big( (2d-4)m_4^4 +(d-2)m_0^2m_2^2 - (2d-4)m_0^2m_3^2  \Big) \omega^4
- (2d-4)m_2^2m_4^2 \omega^2\vec k^2 
+\Big((d-2)m_2^4 -(d-2)m_2^2m_3^2 \Big)\vec k^4 +  \\
+\Big((3d-5)m_2^2m_0^2m_3^2 -(3d-5)m_2^2m_4^4 -(d-1)m_0^2m_2^4 \Big) \omega^2
+ \\
+ \Big( (2d-4)m_2^4m_4^2 -(d-3)m_2^4m_0^2
-(d-3)m_2^2m_4^4 +(d-3)m_3^2m_0^2m_2^2\Big)
\vec k^2 +  \\
+ \Big((d-1)m_2^4m_4^4 -(d-1)m_0^2m_3^2m_2^4 +m_2^6m_0^2\Big) \Big\},  \\
a_{11} = \frac{-2(-\omega^2+\vec k^2+m_1^2)}
{m_1^2(-\omega^2+\vec k^2+m_2^2)(-\omega^2+\frac{m_2^2}{m_1^2}\vec k^2 + m_2^2)}
= -2\frac{-\omega^2+\vec k^2+m_1^2}{D_{vec}D_{gr}},  \\
a_{20}
=-2\frac{((d-2)(m_1^2-2m_4^2)\omega^2+(d-3)m_1^2m_4^2+m_1^2m_2^2-(d-1)m_1^2m_3^2)}
{D},
\\
a_{21}
= \frac{2}{D_{gr}D}\Big(
( m_4^2m_1^2 +m_0^2m_1^2 +2m_3^2m_0^2 -2m_2^2m_0^2 -2m_4^4)\omega^2
+ (m_4^2m_1^2 +m_3^2m_1^2 -m_1^2m_2^2 )\vec k^2 +  \\
+ ( 2m_1^2m_4^4 + m_2^2m_1^2m_0^2 -m_4^2m_1^2m_2^2
-2m_3^2m_0^2m_1^2 )\Big), \\
a_{60} = \frac{-4\omega }{D_{vec}D}\Big(
(m_2^2m_0^2 -2m_3^2m_0^2 -m_4^2m_1^2 +2m_4^4)(d-2)\omega^2
+ ( m_1^2m_2^2 -m_3^2m_1^2 -m_2^2m_4^2 )(d-2)\vec k^2 +  \\
+(-(d-3)m_2^2m_1^2m_0^2+(d-3)m_3^2m_0^2m_1^2-m_2^4m_0^2+(d-1)m_2^2m_0^2m_3^2
+(d-2)m_4^2m_1^2m_2^2 -(d-3)m_1^2m_4^4 -(d-1)m_2^2m_4^4)\Big), \\
a_{61} = \frac{1}{D_{vec}D_{gr}D}\Big(
( 2m_2^2m_0^2 +m_1^4 +4m_4^4 -4m_1^2m_4^2 -4m_0^2m_3^2)(d-2)\omega^4 +\\
+ (4m_1^2m_4^2 +4m_3^2m_0^2 -4m_4^4 -2m_0^2m_2^2 - m_1^4
-2m_3^2m_1^2+m_1^2m_2^2)(d-2)\omega^2\vec k^2 
+ (2m_1^2m_3^2 -m_1^2m_2^2   )(d-2) \vec k^4 +\\
\ \ \ \ \ \ \ \ \
+ (  (4d-8)m_1^2m_2^2m_4^2 -(4d-12)m_2^2m_0^2m_1^2
-(4d-12)m_4^4m_1^2 +(4d-12)m_3^2m_0^2m_1^2
+ \\
+(2d-2)m_2^2m_0^2m_3^2 -2m_0^2m_2^4 -(2d-2)m_2^2m_4^4
-(d-3)m_2^2m_1^4 +(2d-6)m_1^4m_4^2
-(d-1)m_1^4m_3^2+(d-3)m_1^4m_0^2) \omega^2 + \\
+ ( (2d-6)m_1^2m_4^4 +(d-1)m_2^2m_3^2m_1^2 +(d-3)m_0^2m_2^2m_1^2
-(2d-6)m_3^2m_0^2m_1^2-(2d-6)m_4^2m_2^2m_1^2 -m_2^4m_1^2+ \\
+ (2d-4)m_1^4m_3^2-(d-2)m_2^2m_1^4 ) \vec k^2 + \\
+ ( (2d-6)m_1^4m_4^4 -(2d-6)m_2^2m_1^4m_4^2  +(d-3)m_2^2m_1^4m_0^2
+(d-1)m_2^2m_1^4m_3^2
-(2d-6)m_1^4m_3^2m_0^2-m_2^4m_1^4)
\Big),  
\ee
\be
b = -\frac{1}{D}\left[
(   -(d-2)m_1^2 )\omega^4 +
(2d-4)(m_2^2-m_3^2) \omega^2\vec k^2
+ \right. \\\left.+
((d-3)m_2^2m_1^2 +(d-1)m_3^2m_1^2)\omega^2
+ (d-3)( m_3^2m_1^2 -m_1^2m_2^2 )\vec k^2
+ ( m_2^4m_1^2 -(d-1)m_3^2m_2^2m_1^2)\right],  \\
a_{40} = 2\frac{-\omega^2+m_2^2}{D_{vec}},  \\
a_{41} = -\frac{1}{D_{gr}}, 
\\
a_{51} = \frac{-1}{D_{gr}D}\Big(
( m_0^2m_1^2  )\omega^4
+(  2m_4^2m_1^2 +2m_3^2m_0^2-2m_4^4 -2m_0^2m_2^2 ) \omega^2\vec k^2
+ m_1^2(m_3^2 -m_2^2)\vec k^4 + \\
+( m_4^4m_1^2 -m_2^2m_0^2m_1^2 -m_3^2m_0^2m_1^2 )\omega^2
+ (m_0^2m_2^2m_1^2 -2m_4^2m_2^2m_1^2 +m_4^4m_1^2 -m_3^2m_0^2m_1^2)\vec k^2
+ (m_2^2m_0^2m_3^2m_1^2 -m_2^2m_1^2m_4^4   )\Big),  
\\
a_{50} = \frac{2m_1^2
\Big(-m_4^2\omega^2 -(m_3^2-m_2^2)\vec k^2 + m_2^2m_4^2\Big)}{D}, 
\\
c_1 = \frac{-4(d-2)\omega(m_4^2\omega^2+(m_3^2-m_2^2)\vec k^2 -m_2^2m_4^2)}{D},  \\
c_2 = 4\omega\frac{(m_4^4+m_2^2m_0^2-m_3^2m_0^2)}{D},  \\
c_3 = \frac{4\omega}{D_{vec}}
\ee

\subsection{The case of conserved currents:}

For conserved stress-tensor
$k_\mu T^{\mu\nu} = \omega T^{0\nu} + k_iT^{i\mu} = T^{\mu\nu}k_\nu =0$
(\ref{progra}) turns into:
\be
{\cal P}_{\mu\nu,\alpha\beta}T^{\mu\nu}T^{\alpha\beta} 
= 
(a_{10}\omega^2 + a_{20}\omega^2+ b - c_1\omega -a_{60}\omega^3 +a_{61}\omega^4)T_{00}^2
+\\
+ (a_{11}\omega^2+ a_{40} - c_3\omega)T_{0i}^2
+ (a_{21}\omega^2+a_{50}- c_2\omega)T_{00}T_{ii}
+ a_{41}T_{ij}^2 + a_{51}(T_{ii})^2
\label{procons}
\ee
and in $d=4$
\be
\begin{array}{cc}
{\rm tensor\ propagator}: & a_{41} = -\frac{1}{D_2},  \\
{\rm vector\ propagator}: & (a_{11}\omega^2+ a_{40} - c_3\omega) =
-\frac{2}{D_1D_2}\Big(4\omega^2D_{gr} -D_{vec} +(m_1^2-m_2^2)m_2^2\Big)
\end{array}
\ee
No any essential cancelations happen in these expressions, and answers
for the scalar channels look very lengthy and involved.

\subsection{No DVZ jump}

According to our treatment of DVZ discontinuity it occurs
if one removing a ghost from the spectrum via bringing its mass to infinity,
simultaneously removes its contribution from the static potential, the quantity controlled
by $\vec k^2$-dependence. In the Lorentz-invariant case these two things inevitably
happens together. On the contrary, in the non Lorentz invariant case, the
mass of the ghost (=$omega^2$-behaviour) and the static potential are unrelated
and, therefore, the DVZ jump does not happen. Indeed, if one sends $m_0$ to zero,
it makes the ghost mass infinite and removes it from spectrum via canceling the
coefficient in front of $\omega^4$ in $D$, (\ref{Dform}). At the same time, the coefficient
in front of $\vec k^4$ in $D$ becomes non-zero in this case, therefore, not changing
asymptotics of the static potential.

\newpage

\section{Mixing with extra fields and Kaluza-Klein theory
\label{LS}}
\setcounter{equation}{0}

Somewhat amusingly, the mixing of gravity with additional
field was considered already in the seminal paper \cite{PF},
however there it was used just as a technical trick.
Recently this kind of modification of massive gravity was
re-introduced by S.Dubovsky \cite{Du}.
Since then the subject attracts a rapidly increasing
attention.
The idea is to add terms like
\be
h^{\mu\nu}k_\mu \pi_\nu + \pi-{\rm squared\ terms}
\ee
or
\be
h^{\mu\nu}k_\mu k_\nu\pi + \pi-{\rm squared\ terms}
\ee
and their Lorentz-violating analogues to the quadratic
Lagrangian, thus introducing
{\it mixing} of gravity field with something else,
what is denoted by $\pi$ in these formulas.
These $\pi$-fields can be considered as shifted vector or
scalar fields (say, Goldstone fields describing
fluctuations near the vacuum expectation values,
which cause spontaneous violation of gauge and Lorentz
invariances).
There is already convincing evidence that such mixing can
substantially soften the strange properties of massive
gravity and provide healthy {\it perturbatively-reliable}
models with massive graviton and Lorentz violations.

Of course, this conclusion is of no surprise, because
such healthy theory is well known for decades:
this is nothing but the ordinary Kaluza-Klein gravity.

\subsection{Example of Kaluza-Klein graviton,
$d+1=5 \stackrel{{\rm compactification}}{\longrightarrow} d=4$:}

Kaluza-Klein (KK) gravity is ordinary general relativity
in higher $d+m$ dimensional space-time, compactified back into $d$
dimensions.
From $d$-dimensional point of view the theory looks as an
infinite KK tower of fields with different masses, all interacting
among themselves.
However, in quadratic approximation the fields from different
KK sectors (with different masses) do not interact and even mix,
and one can safely consider each sector separately.
In this sector we have a massive $d$-dimensional graviton,
which should be completely free of any kinds of problems,
even gauge invariance (under general coordinate transformations
in $d$-dimensions) is preserved. The question is how this can
be consistent with seemingly unavoidable pathologies of massive
gravity, discussed in the previous sections.
The answer is that this massive graviton is being mixed with the
other KK fields of the same sector (with the same masses),
and this is the simplest possible argument that addition of
extra fields can cure massive gravity from all its potential
problems.

Our task now is to analyze massive KK gravity (i.e. a given
mass level of KK tower) in some detail in order to see
how it works. We restrict consideration to one extra dimension,
compactified on a circle of radius $R_d$, and express everything
in units of $R_d$.
Since even quadratic action for massive KK fields is not
widely known, we begin with its detailed derivation.

\subsubsection{Quadratic part of KK action in a given sector}

Denote different components of $d+1$-dimensional graviton
through $h_{\mu\nu}$,
$A_\mu = h_{\mu d}$ and $\phi = h_{dd}$,
where $\mu,\nu = 0,1,\ldots,d-1$.
The discrete momentum in compactified direction is $k_d=n$,
$n$ is an integer multiple of inverse radius of $R_d$.
We consider a particular sector of a given $n$, it is not mixed
with other sectors in quadratic approximation.

Einstein-Hilbert action in this approximation is
\be
2(kh)_M^2 -k^2h_{MN}^2 - 2(khk)h + k^2h^2 \longrightarrow  \\
\longrightarrow
2(k_\mu h^{\mu\nu} + n A^\nu)^2 + 2(k_\mu A^\mu + n\phi)^2
-(k^2+n^2)(h_{\mu\nu}^2+2A_\mu^2+\phi^2) -  \\
- 2(k_\mu k_\nu h^{\mu\nu} + 2nk_\mu A^\mu + n^2\phi)(h+\phi)
+ (k^2+n^2)(h+\phi)^2 =  \\ =
\Big\{2(kh)_\mu^2 -k^2h_{\mu\nu}^2 - 2(khk)h + k^2h^2\Big\}
+ n^2(h^2-h_{\mu\nu}^2) +  \\
+  2\left(k_\mu k_\nu - k^2\eta_{\mu\nu}\right)A^\mu A^\nu 
+ 4n\left(k_\mu h^{\mu\nu} - k^\nu h\right) A_\nu  +  \\ +
2\left(k^2 h - (khk)\right)\phi
\label{KKac}
\ee
The last line describes $h-\phi$ mixing, which exists even
in massless sector, at $n=0$.

The standard trick in KK theory is to eliminate this mixing
by the shift
\be
h_{\mu\nu} \rightarrow h_{\mu\nu}-\frac{1}{d-2}\phi\eta_{\mu\nu},
\label{n0shift}
\ee
which is taken into account by the standard parametrization
of Kaluza-Klein (KK) metric,
\be
e^{-\frac{\varphi}{d-1}}\left(\begin{array}{cc}
g_{\mu\nu} + e^{\varphi} A_\mu A_\nu & e^{\varphi} A_\mu \\
e^{\varphi} A_\nu & e^{\varphi}
\end{array}\right) =
\left(\begin{array}{cc}
\eta_{\mu\nu} & 0 \\ 0 & 1
\end{array}\right)
+ \left(\begin{array}{cc}
h_{\mu\nu} - \frac{1}{d-1}\varphi\eta_{\mu\nu} & A_\mu \\
A_\nu & \frac{d-2}{d-1}\varphi
\end{array}\right) + {\rm non-linear\ terms}
\ee
with $\phi = \frac{d-2}{d-1}\varphi$.
After this shift we get the quadratic part of KK action in the form:
\be
\Big\{2(kh)_\mu^2 -(k^2+n^2)h_{\mu\nu}^2 - 2(khk)h + (k^2+n^2)h^2\Big\}
+  \\
+  2\left(k_\mu k_\nu - k^2\eta_{\mu\nu}\right)A^\mu A^\nu 
+ 4n\left(k_\mu h^{\mu\nu} - k^\nu h\right) A_\nu  +  \\ +
\frac{d-1}{d-2}\Big(-k^2\phi^2 + 2n\phi (2k_\mu A^\mu - nh)
+ \frac{d}{d-2}n^2\phi^2\Big)
\label{KKac1}
\ee
Now we have a familiar action in massless sector ($n=0$),
which describes gravity plus photodynamics plus additional
neutral Brans-Dicke scalar.
However, for $n\neq 0$ the action is still strange:
graviton has mass $n$ but there is no mass term for the photon
and scalar has mass, different from $n$
(worse than that, the scalar "mass term" has a wrong sign!).
Instead there is severe
mixing between all the three fields:
$h_{\mu\nu}$, $A_\mu$ and $\phi$.
To highlight the problem we can rewrite the last line in
(\ref{KKac1}) as
\be
\frac{d-1}{d-2}\Big(-(k^2+n^2)\phi^2 + 2n\phi (2k_\mu A^\mu - nh)
\Big) + 2\left(\frac{d-1}{d-2}n\phi\right)^2
\ee
and especially strange is the last term.

Of course, diagonalization of this action is not a big problem
-- and in fact  a literal repetition
of that for massless gravity, only in $d+1$ space-time dimensions.
Then, one certainly obtains $(d+1)(d-2)/2$ propagating modes
with the mass $n$ (which form the tensor multiplet of $(d+1)$-dimensional gravity
with the $d$-th component of spatial momentum equal to $n$), $d+1$ zero modes
(corresponding to the Stueckelberg fields) and $d(d+1)/2$ non-propagating modes
(which involve longitudinal graviton).
This result is guessed without any calculations,
after some prejudices are thrown away.
Still, before we proceed to the answer, it is instructive to
analyze immediate peculiarities of KK gravity.

\subsubsection{Properties of massive KK graviton}

\begin{itemize}
\item
The last term in the first line of the $d$-dimensional action
(\ref{KKac}) implies
that the {\bf KK graviton corresponds to the PF choice} $A=B$.

\item
However, {\bf gauge invariance is not broken},
because of the $h-A-\phi$ mixing.
Indeed, one can easily check that (\ref{KKac}) is invariant under
\be
\delta h_{\mu\nu} = k_\mu\xi_\nu + k_\nu\xi_\mu,  \\
\delta A_\mu = n\xi_\mu + k_\mu \zeta, \\
\delta\phi = 2n\zeta
\label{gatra}
\ee

\item
Furthermore, {\bf there is no ghost}, because this is PF gravity.

\item As already mentioned, there should not be any BD instability.

\end{itemize}

\subsubsection{Diagonalizing KK massive sector}

We can now return to the problem of analyzing the
KK Lagrangian (\ref{KKac1}).
In fact, it is better to return one step back --
to (\ref{KKac}).
The peculiarity of KK theory is that the massless ($n=0$)
and massive ($n\neq 0$) should be handled in two
very different ways.
While in the massless sector one makes the celebrated
shift (\ref{n0shift}), which leads to complete
separation of massless graviton, vector and scalar fields
in kinetic matrix, in massive sector one should make
an absolutely different shift:
\be
h_{\mu\nu} \rightarrow h_{\mu\nu} +
\frac{1}{n}(k_\mu A_\nu + k_\nu A_\mu)
- \frac{1}{n^2}k_\mu k_\nu \phi
\label{nshift}
\ee
From the very beginning note the two things:
first, coefficients have $n$ in the denominator, thus this shift
could not be done in the massless sector, and second,
after this shift the new field $h_{\mu\nu}$ is \textit{not}
affected by the gauge transformations (\ref{gatra})
at all -- it should be itself considered as \textit{gauge invariant}.
The result of (\ref{nshift}) on (\ref{KKac}) is spectacular:
the last two lines are fully eliminated, i.e. this shift
reduces the sector $n\neq 0$ to just a single PF massive graviton,
$$
2(kh)_M^2 -k^2h_{MN}^2 - 2(khk)h + k^2h^2 \longrightarrow
$$
$$ \ \stackrel{(\ref{KKac})}{\longrightarrow} \ \Big\{2(kh)_\mu^2
-k^2h_{\mu\nu}^2 - 2(khk)h + k^2h^2\Big\} + n^2(h^2-h_{\mu\nu}^2) +
$$ $$
+  2\left(k_\mu k_\nu - k^2\eta_{\mu\nu}\right)A^\mu A^\nu 
+ 4n\left(k_\mu h^{\mu\nu} - k^\nu h\right) A_\nu  +
2\left(k^2 h - (khk)\right)\phi
$$
$$
\begin{array}{ccc}
n=0 \swarrow (\ref{n0shift}) & \ \ \ \ & (\ref{nshift}) \searrow n\neq 0 \\
&&\\
&&\\
\Big\{2(kh)_\mu^2
-k^2h_{\mu\nu}^2 - 2(khk)h + k^2h^2\Big\} && \Big\{2(kh)_\mu^2
-k^2h_{\mu\nu}^2 - 2(khk)h + k^2h^2\Big\} + n^2(h^2-h_{\mu\nu}^2)
 \\
+ 2\left(k_\mu k_\nu - k^2\eta_{\mu\nu}\right)A^\mu A^\nu
-\left(\frac{d-1}{d-2}\right)k^2\phi^2 &
\end{array}
$$
This is of course what one could expect from the very beginning:
the $\frac{(d+1)(d-2)}{2}$ degrees of freedom of a $d+1$-dimensional
massless graviton can turn either into the
$\frac{(d+1)(d-2)}{2} = \frac{d(d-3)}{2} + (d-2) + 1$ modes
of massless graviton + massless vector + scalar in $d$ dimensions
in the $n=0$ sector
or into the $\frac{(d+1)(d-2)}{2}$ degrees of freedom of
a $d$-dimensional massive graviton in the $n\neq 0$ sectors.
There is simply no room for anything but a massive graviton
in $n\neq 0$ sector, thus nothing like massive vector or
scalar can exist there in addition to massive graviton.

It is also instructive to look at the same counting from the
point of view of the \textit{eigenvalues}.
The field $h_{MN}$ in $d+1$ dimensions had
a kinetic matrix with  $\frac{(d+1)(d+2)}{2}$ eigenvalues
$\lambda$, of which exactly $d+1$ were vanishing.
This left $\frac{(d+1)(d+2)}{2} -(d+1) = \frac{d(d+1)}{2}$
eigenvalues -- as needed for kinetic matrix of a
$d$-dimensional symmetric matrix. We saw in s.4
that all these eigenvalues are indeed non-vanishing --
though not all associated with propagating particles.

\subsubsection{Compactification of a vector field}

To get a better  illustration of what happened in the
previous subsection, one can repeat the same trick for
a photon field: consider a massless $(d+1)$-dimensional
photon $A_M$ and look what happens to it after compactification
to $d$ dimensions, where it turns into a $d$-component
vector $A_\mu$ and a scalar $\phi$.
The Lagrangian
\be
(k_Mk_N- (k^2+n^2)\delta_{MN})A^MA^N =
\Big(k_\mu k_\nu - (k^2+n^2)\Big) A^\mu A^\nu
+ 2n(k_\mu A^\mu) \phi - k^2\phi^2
\ee
In the simplest case of $d=1$ one has in $(1+1)=2$ dimensions just
$(k\phi - nA)^2$ with the kinetic matrix
$$\left(\begin{array}{cc}
- n^2 & kn \\ kn & -k^2
\end{array}\right)$$
and eigenvalues $\lambda = 0$ and $\lambda = k^2+n^2$.
From the one-dimensional point of view, one has instead a single
mode $(k\phi - nA)$ with the eigenvalue $\lambda=1$.
The difference is dictated by normalization:
the usual fact for quadratic forms.

A similar phenomenon occurs for gravitons:
non-vanishing eigenvalues in $d+1$ and $d$
dimensions are in one-to-one correspondence,
but do not literally coincide because of 
different normalizations.

\subsection{DVZ discontinuity}

Since KK graviton is the PF one, it is not a too
big surprise that {\bf DVZ jump occurs} when KK radius
tends to infinity and a graviton mode with a given
$n\neq 0$ becomes massless.
However, in KK case the discontinuity has a simple
explanation: there are additional fields, $A_\mu$ and $\phi$
and they also contribute to the interaction of the
stress tensors.
Discontinuity is exactly the contribution of these extra
fields.

The Born interaction between two stress-tensors
through exchange of KK graviton from the given-$n$ sector
can be immediately read out from the massless case in
(\ref{a5}) by making two changes: $d\rightarrow d+1$
and $k_d \rightarrow n$.
We consider only the interaction between conserved
$d$-dimensional stress tensors, while $T_{\mu d} = T_{dd} = 0$
(actually this does not affect the formulas too much).
The result is:
\be
\frac{1}{k^2+n^2}\Big(
T_{\mu\nu}^2 - \frac{1}{(d+1)\!-2}T^2 \Big)
\label{TTKK}
\ee
and in the massless limit $n\rightarrow 0$
(i.e. $R_d\rightarrow\infty$)
we obtain
\be
\begin{array}{cc}
{\rm massless\ limit\ of\ KK\ gravity\ mode}:
& \ \ \ \ \ \ \ \ \frac{1}{k^2}\Big(
T_{\mu\nu}^2 - \frac{1}{d-1}T^2 \Big)
\end{array}
\label{TTKKmless}
\ee
what is different from the answer for ordinary massless gravity
in $d$ dimensions,
\be
\begin{array}{cc}
{\rm massless\ gravity}: & \ \ \ \ \ \ \ \
\frac{1}{k^2}\Big(
T_{\mu\nu}^2 - \frac{1}{d-2}T^2 \Big)
\end{array}
\label{TTmless1}
\ee
As already said, this is not a big surprise, because
to KK answer the other fields are contributing.
Moreover, (\ref{TTKKmless}) coincides with the contribution
of the massless KK sector with $n=0$, and there the contribution
of the other fields is very simple: at $n=0$ there are no mixings
in (\ref{KKac1}) and the only source of corrections is that
$T_{\mu\nu}$ is coupled to shifted
$h_{\mu\nu} + \frac{1}{d-2}\phi\eta_{\mu\nu}$ instead of $T_{\mu\nu}$.
This provides an additional contribution from the $\phi$-exchange,
which is equal to
\be
+\left(\frac{1}{d-2}\right)^2\frac{d-2}{d-1}\frac{T^2}{k^2} =
\frac{1}{(d-1)(d-2)}\frac{T^2}{k^2}
\ee
(the second factor takes into account the coefficient in front
of kinetic term for $\phi$)
and should be added to the pure graviton exchange (\ref{TTmless1}),
thus changing $-\frac{1}{d-2}$ to
$-\frac{1}{d-2} + \frac{1}{(d-1)(d-2)} = -\frac{1}{d-1}$
which reproduces (\ref{TTKKmless}).

In other words, we observe that in the KK theory the current-current interaction
gets contributions from 3 channels,
\be\label{i1}
\hbox{KK theory}=\hbox{graviton}+\hbox{ghost}+\hbox{scalar field}
\ee
while in the massless gravity these are 2 channels
\be\label{i2}
\hbox{Massless gravity}=\hbox{graviton}+\hbox{ghost}
\ee
and in the PF theory, where the ghost is removed out of the spectrum by taking its mass
to infinity, there is only the graviton contribution:
\be\label{i3}
\hbox{PF theory}=\hbox{graviton}
\ee
As we saw above, the contributions of the ghost and the scalar field exactly
cancel each other so that the current-current interactions in the KK theory and in
the PF theory coincide. If, however, one considers a KK theory with several compactified
dimensions, i.e. with several scalar fields added, this compensation will no longer
take place, and all the three interactions, (\ref{i1}), (\ref{i2}) and (\ref{i3})
will be different.

\newpage

\section{Massive gravity
within and beyond the quadratic approximation}
\setcounter{equation}{0}

In this section we make very brief comments about the other
aspects of massive gravity, some of which are not directly seen at
linearized level, but are in fact direct
consequences of the properties of quadratic theory.
These are pronounced and affective phenomena, but do not add
much new to the theoretical aspects of the problem.

\subsection{Vainshtein radius \cite{Va}}

This celebrated result is often considered as the clear proof
of pathology of PF massive gravity, though the actual statement
is much simpler.
The new modes, revived by addition of mass terms to Einstein-Hilbert
action, get kinetic terms because at least some of them are
Stueckelberg fields.
We emphasize once again that one does
not modify the kinetic part of Einstein-Hilbert action,
only masses (more generally, a non-trivial potential) are added.
This means that the new kinetic terms enter with small coefficients
$m^2$, $m^2k^2s^2$, and this means that the physical fields are
$ms$ rather than $s$. This means that higher-degree terms, say,
$m^2s^3 = \frac{1}{m}(ms)^3$ are actually entering with
{\it large} couplings $\sim m^{-1}$, and that the theory
is actually strongly coupled, moreover interactions get stronger
with decrease of $m^2$.
Interaction effects are of course falling with the distance,
therefore very far away, beyond some "Vainshtein radius"
perturbative regime can still be reliable, but in general
perturbation theory does not make much sense at small and
even moderate distances.

This strong coupling effect is a result of explicit breakdown of
gauge symmetry and it is rather similar to what happens when
Higgs mass goes beyond the unitary limit in the case of spontaneously broken
gauge symmetry in Yang-Mills theory.
Nothing like this strong-coupling phenomena occurs in the
case of gauge invariant Kaluza-Klein massive gravity and hence
it can be avoided when gravity mixes with other fields.

\subsection{Ghosts}

Ghost are the fields which enter Lagrangian with the wrong sign
in front of the time-derivative term $\dot\phi^2$.
What happens to them is that they grow in time until this
growth is stopped by non-linear terms in the action or the
theory gets into a different phase with another spectrum of
quasiparticles. In this sense this is a phenomenon of the same
class as the previous one: ghosts make perturbative treatment
unreliable and most often inadequate.
In this context one often speaks about "negative norms"
and "violation of unitarity" -- but this actually refers
to "perturbative unitarity", implying that the actual spectrum
of the theory is more-or-less accurately described by quadratic
part of the Lagrangian, what is not always the case.
While occurrence of ghosts clearly means that the theory is
not what we thought it will be, it does not obligatory means
that it is ill and un-curable -- it is just not perturbative
and most probably strongly coupled.

There are three ways to deal with ghosts.
The first option is to study the theory
as it is (what is rarely done, see, however,
\cite{AS} or
the series of works about the "pathological" Hamiltonian $H=xp$
\cite{xp}).
The second option is
to say that the growth of ghosts is slow enough to be acceptable
(e.g., the ghost formally appears in the Lorentz invariant
massive gravity with $A=2B$, which corresponds to addition of
the cosmological term to the Einstein-Hilbert Lagrangian, what
happens is that the fields grow together with the growth
of the Universe itself).
The third, and most popular option is to fine-tune parameters
in order to "eliminate" ghosts, for example, to give them an
infinite mass.

\subsection{Boulware-Deser mode \cite{BD,RT} }

The last option is exactly the one, "distinguishing" the Pauli-Fierz
gravity with $A=B$. However, the way in which the ghost acquires
infinite mass is somewhat special and actually unreliable.
The mass is infinite because the coefficient in front of kinetic
term vanishes! As we already discussed, this is what actually
makes the theory strongly coupled, moreover, this of course makes
the fine-tuning fully unreliable. Any minor deformation of
the theory destroys the fine-tuning and brings the ghost back
to existence. Boulware-Deser instability is a concrete example
of this phenomenon: switching on non-trivial background metric
contributes to the coefficient in front of the kinetic term
and shifts it away from zero. Since background is arbitrary
it can not be compensated by variation of just two adjusting
constants $A$ and $B$.

Of course, nothing like this happens in ghost-free generalizations
of PF gravity such as KK or other models
with extra fields.
Ghosts are absorbed into these additional fields in a universal,
background independent way.

\subsection{DVZ discontinuities}

As we already explained these discontinuities occur when
comparing two theories with different sets of fields and thus are
of no surprise.
However, from the point of view of concrete example of massless
gravity, it is instructive to distinguish between two situations:
when one compares it to theory with {\it more} fields and with
{\it less} fields.

The first is, for example, the case of Kaluza-Klein gravity:
in a given mass sector of Kaluza-Klein theory there are two
{\it more} fields, $A_\mu$ and $\phi$, which mix with the massive
graviton and also contribute to Newton-like interaction.
This additional contribution explains the difference.

The second is the Pauli-Fierz gravity, where one of the modes,
contributing to Newton interaction in massless theory, becomes
a ghost and is thrown away "by hands".
This explains the discontinuity. It is in no way a property of
massive gravity, it is a property of this artificial
throw-away prescription.
The reason why the would-be ghost is allowed in massless gravity
is that it has the same mass as the other (non-ghost) degrees
of freedom, and this results into decoupling of the negative norm states
from the spectrum which
leaves the theory perturbatively consistent (we do not
speak about UV problems of quantum gravity here).
Notably, the same kind of absorption could be expected if instead of
the Pauli-Fierz choice $A=B$ we put $A=2B$ when the two masses
continue to coincide, $M=m$. Then there will be no DVZ discontinuity.
This is the case of the cosmological term added to the Einstein-Hilbert
action, and, because of presence of linear terms in the Lagrangian,
one has to re-expand this latter above the non-flat AdS vacuum where the linear
term cancels \cite{AdS}
(see also footnote 7 in \cite{RT} about this option).

\subsection{Tachyons}

The name tachyon refers to superluminal propagation.
However, in modern literature it is actually used to mean something
different: a mode, that reflects perturbative instability of the
background around which the perturbative theory is developed.
Technically this means that there is a pole in the propagator at
vanishing frequency, for example, $M^2<0$.
Physically this means that a phase transition of the first kind
occurs from perturbatively unstable vacuum to another one,
stable at least perturbatively (false vacua which are separated
from the real ones by potential barriers do not have tachyons in
perturbative spectra, their instability is essentially
non-perturbative phenomenon).
Such phase transition takes place spontaneously and independently
in all point of the space: this can look like a propagation of
non-causal (superluminal) particle, but has a clear reason,
and processes taking place in different places are indeed casually
unrelated.
Tachyons are in no way a problem of the theory -- they signal just
that we treat (interpret) it in a wrong way.
It is of course not always easy to find the write vacuum, this
is often related to finding the write non-perturbative formulation
of the theory. The celebrated example of such lasting study was
interpretation of tachyons in string theory (for open superstring
the puzzle was partly resolved by A.Sen in \cite{Sen}, in general
string models it remains obscure).

What is the right vacuum
of Lorentz-violating massive gravity with tachyons, and what at all is its
adequate non-perturbative formulation is an open problem
(not surprisingly given by the young age of this subject).
However, one may expect this vacuum has not to be homogeneous,
due to non-trivial dispersion laws (see s.5.6). Moreover, phenomenologically
this may be not that bad, since the scale of these inhomogeneities is
determined by inverse masses that breaks Lorentz invariance, i.e. which are
very small. Inhomogeneities at such large scales are quite possible.

\subsection{Superluminal propagation \cite{Dsl,OR}}

We reserve the word superluminal propagation for phenomenon
which is (at least looks) different from tachyons, i.e. is
not related in any clear way to perturbative instabilities
and phase transitions. This is occurrence of poles in the
propagator at $\omega = ck$ with $c>1$.
Such poles are forbidden by Lorentz-invariance, but
we saw that they naturally appear in the spectrum of
massive gravity when Lorentz invariance is violated
(for example, $c^2=m_1^2/m_2^2$).
The meaning of superluminal dispersion laws remains
controversial, we feel that most people find them unpleasant,
referring mostly to causality arguments, which are also
used against time-machines -- which surely are not
forbidden by the laws of nature (see \cite{mtm} for the
most recent discussion).
Quite similarly, superluminals are unavoidably present
in our theories, whether we like them or not.
Whenever one perturbs a light-like dispersion law
$\omega^=k^2$ for a collection of light-like particles
(two polarizations are already sufficient) in a
Lorentz-violating way,
the eigenvalues split, and one goes above, another below
$c=1$: this is a fundamental law of linear algebra, well
known in the level-splitting theory in quantum mechanics.
It is important that Lorentz violation need not be
"fundamental" (i.e. explicitly written down into Lagrangian):
a Lorentz-violating background is already enough.
The celebrated example is emergency of
superluminal photons in curved space \cite{Dsl},
where the effective Lagrangian acquires a quantum correction
(1-loop of fermion or any other field in gravitational
background) and becomes
\be
F_{\mu\nu} F_{\alpha\beta} \Big(g^{\mu\alpha}g^{\nu\beta} +
{\rm const}\cdot R^{\mu\alpha\nu\beta} + \ldots\Big)
\label{DSL}
\ee
what -- according to the above-mentioned linear-algebra
theorem -- unavoidably leads to dispersion law with $c>1$
for one of photon polarizations.

Implications of superluminals are still badly understood.
It is not even fully clear if they can be used to construct
time-machines, either big
(of astrophysical scale) \cite{tm}
or mini (of Planckian scale) \cite{mtm}. In any case, we repeat
that time machines are allowed already in ordinary
General Relativity and,
within hypothetical TeV-gravity models \cite{TeVgr},
might even be massively created (though immediately evaporate)
in particle collisions in modern accelerators \cite{mtm}.

Most important, our intuition about Lorentz-non-invariant
theories is still very underdeveloped. Always we discuss
our models (in this context) from the point of view of some
outside observer which believes into {\it fundamental}
Lorentz invariance: we ask what happens if we look on a
superluminal from another frame (what happens is a singularity
in dispersion rule $\omega + \beta k = c(k+\beta \omega)$
i.e. $\omega = \frac{c-\beta}{1-\beta c}$
at $\beta = 1/c$ for $c>1$ instead of the usual zero at $\beta = c$
for $c<1$) and discuss whether this good or bad.
However, if we just look at a theory {\it per se}, it is absolutely
unclear what is going to be bad about it: there simply is no
Lorentz invariance and it is unclear why at all one should ask
what happens if you make such a transformation.
It is like writing down a theory with rescaled $x$-coordinate
and apply Lorentz transformation with and old (not-rescaled) $x$.
Even more important would be to examine what can be wrong with
just a combination of two fields,
\be
L = \frac{1}{2}(\dot \phi_1^2 - c_1^2 \vec\nabla \phi_1^2 +
\dot \phi_2^2 - c_2^2 \vec\nabla \phi_2^2) + V(\phi_1,\phi_2)
\ee
If $c_2>c_1$ the second field is a superluminal from the point
of view of the first one, but nothing wrong is expected if we look
on the first field from the point of view of the second.

\subsection{Radiation of massive gravitons}

As mentioned in the introduction, one of our original questions was if the
"pathologies" of massive gravity can somehow affect the analysis
\cite{grarad} of massive-graviton radiation in TeV-gravity models,
a worry, naturally implied by the original analysis in \cite{New}.
However, the Kaluza-Klein theory
(which is in the base of the Tev-gravity models) seems to be free
of any problems, true or imaginary, of generic massive gravity
and our current feeling is that one can treat radiation of massive
Kaluza-Klein gravitons in the straightforward and naive way,
as suggested \cite{grarad}. The problem, however, deserves an
independent analysis.

\section*{Acknowledgements}

We are indebted for hospitality and support to Prof.T.Tomaras and
the Institute of Theoretical and Computational Physics of
University of Crete during the summer of 2008, where this work was done.
We are specially grateful to T.Tomaras for interest, long discussions and collaboration.

Our work is partly supported by Russian Federal Nuclear Energy
Agency, by the joint grants 09-02-91005-ANF and 09-01-92440-CE,
by the Russian President's Grants of
Support for the Scientific Schools NSh-3035.2008.2
(A.Mir.,Al.Mor.) and NSh-3036.2008.2 (S.Mir.,An.Mor.), by RFBR grants
07-02-00878 (A.Mir.), 08-02-00287 (S.Mir.), 07-02-00645
(Al.Mor.) and 07-01-00526 (An.Mor.).



\begin{thebibliography}{12}

\bibitem{F} M.Fierz, Helv.Phys.Acta {\bf 12} (1939) 3

\bibitem{PF}
M.Fierz and W.Pauli,
Proc.Roy.Soc. {\bf 173} (1939) 211

\bibitem{magr} A.Logunov, {\sl Relativistic Theory of Gravity,} Commack, USA: Nova Sci.
Publ. (1998) 114 p.\\
G.t'Hooft, arXiv:0708.3184 [hep-th]

\bibitem{MGatt}
G.Dvali, G.Gabadadze and M.Porrati,
Phys.Lett. {\bf B485} (2000) 208, hep-th/0005016\\
N.Arkani-Hamed, H.Georgi and M.D.Schwartz, Ann.Phys. {\bf 305} (2003) 96 (hep-th/0210184)\\
A.Lue,
Phys.Rev. {\bf D66} (2002) 043509,
hep-th/0111168\\
M.Porrati, JHEP {\bf 04} (2002) 058 (hep-th/0112166)\\
A.Gruzinov, astro-ph/0112246

\bibitem{othex} J.Bekenstein, Phys.Lett. {\bf B202} (1988) 497; Phys.Rev. {\bf D70} (2004) 083509,
astro-ph/0403694; PoS {\bf JHW2004} (2005) 012, astro-ph/0412652

\bibitem{MT1} A.Mironov, S.Mironov, A.Morozov and And.Morozov,
arXiv:0910.5245 (hep-th)

\bibitem{DVZ} H.van Dam and M.Veltman, Nucl.Phys. {\bf B22}
(1970) 397; \\
V.Zakharov, JETP Lett. {\bf 12} (1970) 312

\bibitem{Va} A.Vainshtein, Phys.Lett. {\bf 39B} (1972) 393

\bibitem{BD} D.G.Boulware and S.Deser, Phys.Rev. {\bf D4} (1972) 3368

\bibitem{Ru} V.Rubakov, hep-th/0407104

\bibitem{Du} S.Dubovsky, JHEP {\bf 0410} (2004) 076 (hep-th/0409124)

\bibitem{TiT} S.Dubovsky, P.Tinyakov and I.Tkachev, Phys.Rev.Lett. 94 (2005) 181102 (hep-th/0411158);
Phys.Rev. D72 (2005) 084011 (hep-th/0504067)

\bibitem{RT} V.Rubakov and P.Tinyakov, Phys.Usp. {\bf 51} (2008) 759-792, arXiv:0802.4379

\bibitem{grapa} G.Dvali, O.Pujolas, M.Redi, Phys.Rev.Lett. {\bf 101} (2008) 171303, arXiv: 0806.3762

\bibitem{Ni} P.van Nieuwenhuizen,
Nucl.Phys. {\bf B60}
(1973) 478-492

\bibitem{KM} Ya.Kogan and A.Morozov,
JETP {\bf 61} (1985) 1-8 (ZhETF {\bf 88} (1985) 3-16)

\bibitem{3dph} S.Deser, R.Jackiw and S.Templeton,
Phys.Rev.Lett. {\bf 37B} (1971) 95

\bibitem{New} P. Van Nieuwenhuizen, Phys.Rev. {\bf D7} (1973) 2300-2308

\bibitem{grarad} A.Mironov, A.Morozov,
Pisma ZhETF {\bf 85} (2007)
9-14 (JETP letters {\bf 85} (2007) 6-11), hep-ph/0612074

\bibitem{TeVgr} N.Arkani-Hamed, S.Dimopoulos and G.Dvali,
Phys.Lett.B429 (1998) 263-272, hep-ph/9803315(1998);
Phys.Today 55N2 (2002) 35-40

\bibitem{mbh} T.Banks and W.Fishler, hep-th/9906038\\
S.B.Giddings and S.Thomas, Phys.Rev. {\bf D65} (2002) 056010\\
S.Dimopoulos and G.Landsberg, Phys.Rev.Lett. {\bf 87} (2001) 161602\\
{\bf References and a review can be found in:}\\
P.Kanti, Int.J.Mod.Phys. {A19} (2004) 4899\\
G.Landsberg, J.Phys. {\bf G32} (2006) R337, hep-ph/0607297\\
M.Cavaglia, R.Godang, L.Cremaldi and D.Summers, hep-ph/0609001\\
{\bf For mini-black-holes in cosmic ray events, see:}\\
A.Mironov, A.Morozov and T.Tomaras,
Sov.J.Nucl.Phys. (Yad.Fiz.), hep-ph/0311318\\
A.Cafarella, C.Coriano and T.Tomaras, JHEP {\bf 0506} (2005) 065,
hep-th/0410358\\
{\bf For mini-black-holes in neutrino experiments, see a very
detailed review and references 
in:}\\
L.Anchordoqui, T.Paul, S.Reucroft and J.Swain, Int.J.Mod.Phys. {\bf
A18}
(2003) 2229

\bibitem{mtm} I.Ya.Aref'eva, I.V.Volovich,
arXiv:0710.2696\\
A.Mironov, A.Morozov and T.Tomaras,
Facta Univ.Ser.Phys.Chem.Tech.4 (2006) 381-404, arXiv:0710.3395

\bibitem{OR} M.Osipov and V.Rubakov, Class.Quant.Grav. {\bf 25} (2008) 235006, arXiv:0805.1149

\bibitem{Por} C.Deffayet, G.Dvali, G.Gabadadze and A.Vainshtein,
hep-th/0106001\\
M.Porrati, 
hep-th/0203014 v2

\bibitem{UFN2} A.Morozov,
Sov. Phys. Usp. {\bf 35} (1992) 671-714

\bibitem{Dsl} {\bf Superluminal propagation in general relativity
was discussed in:}\\
A.D.Dolgov and I.B.Khriplovich, Phys.Lett. {\bf A243} (1998) 117, hep-th/9708056\\
A.D. Dolgov and I.D. Novikov, Phys.Lett. {\bf B442} (1998) 82-89, gr-qc/9807067\\
S.Liberati, S.Sonego and M.Visser, Annals Phys. {\bf 298} (2002) 167, gr-qc/0107091\\
{\bf For spectacular fresh analysis of this subject see the
recent papers:}\\
T.J.Hollowood and G.M.Shore, Phys.Lett. {\bf B655} (2007) 67, arXiv:0707.2302;
Nucl.Phys. {\bf B795} (2008) 138, arXiv:0707.2303; JHEP {\bf 0812} (2008) 091,
arXiv:0806.1019

\bibitem{KS} V.A.Kostelecky and S.Samuel, Phys.Rev.{\bf D39}
(1989) 683

\bibitem{LVm} J.Bjorken, Ann.Phys. {\bf 24} (1963) 174, see also hep-th/0111196\\
S.Coleman, S.Glashow, Phys.Rev. {\bf D59} (1999) 116008 hep-ph/9812418

\bibitem{LV} V.A.Kostelecky and R.Potting, Nucl.Phys. {\bf B359}
(1991) 545\\
D.Colladay and V.A.Kostelecky, Phys.Rev. {\bf D55}
(1997) 6760; Phys.Rev. {\bf D58} (1998) 116002\\
R.~Bluhm, Lect. Notes Phys. {\bf 702} (2006)191,  hep-ph/0506054;
Int. J. Mod. Phys. D {\bf 16} (2008) 2357, hep-th/0607127;
arXiv: 0801.0141\\
D.~Colladay, AIP Conf. Proc. {\bf 672} (2003) 65,  hep-ph/0301223\\
V.~A.~Kostelecky and N.~Russell,
arXiv: 0801.0287\\
O.~Bertolami and J.~Paramos, Phys. Rev. D {\bf 72} (2005) 044001,
hep-th/0504215\\
R.~Bluhm and V.~A.~Kostelecky,
Phys. Rev. D {\bf 71} (2005) 065008,  hep-th/0412320\\
M.~Gomes, T.~Mariz, J.~R.~Nascimento and A.~J.~da~Silva,
Phys. Rev. D {\bf 77} (2008) 105002, arXiv: 0709.2904\\
R.~Bluhm, S.~H.~Fung and V.~A.~Kostelecky,
Phys. Rev. D {\bf 77} (2008) 065020, arXiv: 0712.4119\\
S.~M.~Carroll, {\it Aether compactification},
arXiv: 0802.0521\\
R.~Bluhm, N.~L.~Gagne, R.~Potting and A.~Vrublevskis,
Phys. Rev. D {\bf 77} (2008) 125007, arXiv: 0802.4071\\
R.~Obousy and G.~Cleaver,
arXiv: 0805.0019\\
J.~W.~Moffat, Int. J. Mod. Phys. D
{\bf 12} (2003) 1279 , hep-th/0211167\\
O. Bertolami, R. Lehnert, R. Potting, A. Ribeiro,
Phys. Rev. D {\bf 69} (2004) 083513, astro-ph/0310344\\
S.~M.~Carroll and E.~A.~Lim,
Phys. Rev. D {\bf 70} (2004) 123525, hep-th/0407149\\
P.Ferreira, B.Gripaios, R.Saffari
and T.Zlosnik, Phys. Rev. D {\bf 75} (2007) 044014, astro-ph/0610125\\
Arianto, F.~P.~Zen, B.~E.~Gunara, Tryanta and Supard,
JHEP {\bf 09} (2007) 048,  arXiv: 0709.3688\\
J.~W.~Moffat and V.~T.~Toth,
arXiv: 0710.0364\\
L.~Grisa, 
arXiv: 0803.1137\\
R.~Obousy and G.~Cleaver,
arXiv:0805.0019\\
T.Mariz, J.Nascimento, A.Petrov, A.Santos and A.da Silva,
arXiv:0807.4999 \\
Z.Berezhiani and O.Kancheli, 
arXiv:0808.3181\\
E.Kiritsis and V.Niarchos, arXiv:0808.3410\\
P.Koroteev and M.Libanov, arXiv:0901.4347\\
M.Visser, arXiv:0902.0590

\bibitem{LBO} L.Okun, {\sl Leptons and Quarks}, NorthHolland, Amsterdam, 1982

\bibitem{AS} A.Smilga,
Phys.Lett. {\bf B632} (2006) 433-438, hep-th/0503213

\bibitem{xp} M. V. Berry and J. P. Keating,
SIAM Review {\bf 41 (2)} (1999) 236\\
G.Sierra, Nucl.Phys. {\bf B776} (2007) 327-364, math-ph/0702034\\
G.Sierra and P.Townsend, Phys.Rev.Lett. {\bf 101} (2008) 110201, arXiv:0805.4079

\bibitem{AdS} 
I.I. Kogan, S. Mouslopoulos, A. Papazoglou and L. Pilo. Nucl. Phys. {\bf B625}
(2002) 179, hep-th/0105255; Phys. Lett. {\bf B503} (2001) 173, hep-th/0011138\\
M. Porrati. Phys. Lett. {\bf B498} (2001) 92, hep-th/0011152\\
A. Karch, E. Katz and L. Randall. JHEP {\bf 0112} (2001) 016, hep-th/0106261\\
P.A.Grassi and P. van Nieuwenhuizen, Phys.Lett. {\bf B499} (2001) 174-178 0011278\\
Y.S.Myung, hep-th/0012082\\
{\bf For other aspects and references see}\\
M.Novello and R.P.Neves, Class.Quantum Grav. {\bf 20} (2003) L67-L73

\bibitem{Sen} A.Sen, JHEP {\bf 0204} (2002) 048, arXiv:hep-th/0203211

\bibitem{tm} See the detailed bibliography in the second paper of ref.\cite{mtm}

\bibitem{hder} A.Morozov,
Theor.Math.Phys. {\bf 157} (2008) 1542-1549,
arXiv: 0712.0946 v3; \\
P.Dunin-Barkovsky and A.Sleptsov,
Theor.Math.Phys., arXiv:0801.4293

\bibitem{Stue}
E.Stueckelberg,
Helv. Phys. Acta 11, 225-244,
Helv. Phys. Acta 11, 299-312, Helv. Phys. Acta 11, 312-328


\end{thebibliography}
\end{document}